\documentclass[11pt,a4paper]{article}
\usepackage{jheppub}

\usepackage{amsmath,amsthm, amssymb,epsfig,euscript,array,xcolor,dsfont,comment, physics}

\usepackage{pdfpages}
\usepackage[normalem]{ulem}

\definecolor{green}{rgb}{0.0, 0.5, 0.0}
\definecolor{MC}{HTML}{b39b00}
\definecolor{LB}{HTML}{bf6b00}
\definecolor{SB}{HTML}{8900dc}
\definecolor{AH}{HTML}{019c79}
\definecolor{RK}{HTML}{0082dc}
\definecolor{violet}{rgb}{0.58, 0.0, 0.83}

\newcommand{\vol}{{\cal V}}
\newcommand{\ap}{\alpha'}

\newcommand{\gs}{g_s}
\newcommand{\K}{K\"ahler~}
\newcommand{\A}{{\cal A}}
\newcommand{\B}{{\cal B}}

\newcommand{\D}{{\cal D}}

\renewcommand{\O}{{\cal O}}
\let\latexd\d
\renewcommand{\dd}{\text{d}}
\newcommand{\cloop}{c_{\text{loop}}}
\newcommand{\Kahler}{K\"ahler~}
\newcommand{\V}{\mathcal{V}}
\newcommand{\cor}{\textit}

\def\KK{{\scriptscriptstyle KK}}
\def\W{{\scriptscriptstyle W}}
\def\SM{{\scriptscriptstyle SM}}
\def\CW{{\scriptscriptstyle CW}}

\title{Loop Blow-up Inflation}

\author[1,2,3]{Suk\latexd{r}ti Bansal,}
\author[4,5]{Luca Brunelli,}
\author[4,5]{Michele Cicoli,}
\author[3]{\\Arthur Hebecker,}
\author[3]{Ruben Kuespert}

\affiliation[1]{\footnotesize Institut f\"ur Physik und IRIS Adlershof, Humboldt Universit\"at zu Berlin,
Zum Großen Windkanal 2, 12489 Berlin, Germany}
\affiliation[2]{\footnotesize Institute for Theoretical Physics, Karlsruhe Institute of Technology, Wolfgang-Gaede-Straße 1, 76131 Karlsruhe, Germany}
\affiliation[3]{\footnotesize Institut f\"ur theoretische Physik, Universit\"at Heidelberg, Philosophenweg 19, 69120 Heidelberg, Germany}
\affiliation[4]{\footnotesize Dipartimento di Fisica e Astronomia, Universit\`a di Bologna, via Irnerio 46, 40126 Bologna, Italy}
\affiliation[5]{\footnotesize INFN, Sezione di Bologna, viale Berti Pichat 6/2, 40127 Bologna, Italy}

\footnotesize\emailAdd{bansalsu@physik.hu-berlin.de} 
\footnotesize\emailAdd{l.brunelli@unibo.it}
\footnotesize\emailAdd{michele.cicoli@unibo.it}
\footnotesize\emailAdd{a.hebecker@thphys.uni-heidelberg.de} 
\footnotesize\emailAdd{r.kuespert@thphys.uni-heidelberg.de}

\abstract{We present a new model of string inflation driven by a blow-up K\"ahler modulus of type IIb compactifications with a potential generated by string loops. Slow-roll is naturally realized thanks to the fact that the blow-up mode is a leading-order flat direction lifted by string loops which are unavoidable and generate a plateau at large field values. We check that throughout the whole inflationary dynamics the effective field theory is under control. We perform a phenomenological analysis determining the exact number of efoldings by studying the post-inflationary evolution. We determine the values of the microscopic parameters which lead to agreement with CMB data, together with the prediction of a tensor-to-scalar ratio of order $r\sim 10^{-5}$.} 


\begin{document}

\maketitle

\bigskip

\section{Introduction}
\label{sc:intro}

Slow-roll inflation requires a scalar potential with a sufficiently flat region, as quantified by the smallness of the slow-roll parameters. While it is not hard to find potentials with this feature in effective quantum field theory \cite{Lyth:1998xn, Lyth:2007qh, Lyth:2009zz, Martin:2013tda}, realizing them in the 4d effective field theory (EFT) derived from string theory is challenging~\cite{Linde:2005dd, McAllister:2007bg, Baumann:2014nda, Cicoli:2011zz, Burgess:2013sla, Cicoli:2023opf}. This may be related to the well-known difficulties of realizing de Sitter space in string theory. 

However, as pointed out in \cite{Cicoli:2011zz, Burgess:2013sla, Burgess:2014tja, Burgess:2016owb, Cicoli:2023opf}, flat potentials arise rather naturally in the K\"ahler moduli sector of type IIb flux compactifications, due to the existence of approximate rescaling shift symmetries for all moduli orthogonal to the overall volume. In this framework, the volume can be stabilized at a sufficiently large value and an appropriate uplift to an almost-Minkowski vacuum can be realized. Assuming that the challenge of a realistic large volume compactification has been met, a suitable inflationary plateau can appear in many cases \cite{Conlon:2005jm, Bond:2006nc, Cicoli:2008gp, Cicoli:2016chb, Broy:2015zba, Cicoli:2016xae,Cicoli:2017axo,Cicoli:2020bao, Cicoli:2011ct, Lust:2013kt, Blumenhagen:2012ue}.\footnote{For constructions of inflection point inflation relying on the volume modulus as the inflaton see~e.g.~\mbox{\cite{Linde:2007jn,Conlon:2008cj,Cicoli:2015wja, Antoniadis:2020stf,Antoniadis:2021lhi,Bera:2024zsk}}.}
In this paper we derive a new and particularly simple model within this general class of constructions: our inflaton is a blow-up mode whose potential is generated by string loop corrections to the K\"ahler potential.

We recall that, due to the no-scale structure of type IIb flux compactifications, the naively dominant $1/{\cal V}^2$ terms in the scalar potential cancel~\cite{Giddings:2001yu}. As a result, the leading-order K\"ahler moduli scalar potential is generated by $\mathcal{O}(\alpha'^3)$ effects and scales like \cite{Becker:2002nn}
\begin{equation}
	V \sim \frac{|W_0|^2}{{\cal V}^3}\,,
 \label{Vleading}
\end{equation}
where $W_0$ is the constant superpotential generated by fluxes and ${\cal V}$ is the volume modulus. More specifically, we assume that we are in the regime of validity of the Large Volume Scenario (LVS)~\cite{Balasubramanian:2005zx, Conlon:2005ki}. This implies that the total Calabi-Yau volume takes the form $\V=\tilde{\V}(\tau_I)-\tau_s^{3/2}$, with 4-cycle K\"ahler moduli $\{\tau_I\}=\{ \tau_0,\dots,\tau_n\}$ and $\tau_s$. We may eliminate one of the 4-cycle variables in favor of the total volume $\V$ such that the scalar potential takes the form $V=V({\cal V},\tau_i,\tau_s)$, where now $\{ \tau_i\}=\{ \tau_1,\dots,\tau_n\}$. We will always be in the regime $\tau_s^{3/2}\ll \V$, referring to $\V$ as the volume, to $\tau_s$ as the small-cycle modulus, and to the $\{\tau_i\}$ as the `additional K\"ahler moduli.' A key-result of the LVS proposal \cite{Cicoli:2008va} is that, under rather general assumptions, $\tau_s$ and ${\cal V}$ get stabilized while, in the region $\tau_s^{3/2}\ll \tau_i^{3/2} \lesssim \V$, the potential for the $\tau_i$, $i=1,\dots, n$, remains flat at leading order: $V({\cal V},\tau_i,\tau_s)=V({\cal V},\tau_s)$. 

At subleading order, perturbative and non-perturbative $\tau_i$-dependent corrections do in general arise. When the $\tau_i$ take large values, non-perturbative effects of the form $e^{-2\pi\tau_i/N_i}$ are suppressed by $\tau_i\gg 1$. Another source for the potential of the $\tau_i$ moduli are perturbative corrections to the K\"ahler potential. However, as studied in \cite{vonGersdorff:2005bf, Berg:2005ja, Berg:2005yu, Berg:2007wt,Cicoli:2007xp, Cicoli:2008va, Gao:2022uop}, string loops enjoy an extended no-scale structure which makes them in general subleading with respect to (\ref{Vleading}). Additional perturbative corrections arise from higher F-term $\mathcal{O}(\alpha'^3)$ corrections but these come along with a larger suppression power of $\mathcal{V}$ \cite{Ciupke:2015msa}.

The leading-order flatness described above has allowed for the construction of several inflationary scenarios for the simplest situation with a single flat direction: the original model of \cite{Conlon:2005jm, Bond:2006nc} where inflation is driven by a blow-up mode lifted by non-perturbative corrections, fibre inflation models \cite{Cicoli:2008gp, Broy:2015zba, Cicoli:2016chb,Cicoli:2016xae,Cicoli:2017axo,Cicoli:2020bao} where the inflaton is a bulk fibre modulus with a potential generated by string loops or higher $\alpha'^3$ effects, and the models of \cite{Cicoli:2011ct, Lust:2013kt, Blumenhagen:2012ue} where the inflationary potential is generated by poly-instanton effects and the inflaton is either a bulk fibre divisor or a rigid 4-cycle with Wilson lines. 

As noted in \cite{Cicoli:2008gp}, loop corrections represent a potential problem for the original\linebreak model~\cite{Conlon:2005jm} since they would destroy slow-roll. We will estimate that this is indeed the case if the coefficient of these loop corrections is at least of order $10^{-6}$. Note that inflaton-dependent open string loops could be absent by brane construction but, as pointed out in~\cite{Gao:2022uop}, inflaton-dependent closed string loops are unavoidable, unless the effective field theory in the region close to the inflaton 4-cycle is effectively $\mathcal{N}=2$ (a situation which is hard to envisage without eliminating the non-perturbative superpotential term). In this paper we will however show that, going to larger field values which are still well within the K\"ahler cone, string loops become sufficiently suppressed and can naturally lead to slow-roll. We note that such a possibility was pointed out in \cite{Cicoli:2011zz}, but has not been analyzed.

Our present proposal builds on the following key observation: consider the situation with a single leading-order flat direction $\tau_\phi$ which we assume to be a blow-up mode. Then, very generically and without any particular assumption about the functional form of loop corrections, the LVS setting allows for inflation in a regime where $\tau_\phi\lesssim {\cal V}^{2/3}$. To see this, let us disregard the stabilized modulus $\tau_s$, treat the volume as a fixed parameter, ${\cal V}\gg 1$, and write the loop-corrected potential as:
\begin{equation}
\label{eq:potential-prediction}
	V \sim \frac{|W_0|^2}{{\cal V}^3} \left[{\cal O}(1)- 	\frac{c_\text{loop}}{{\cal V}^{1/3}}\, f\left(\frac{{\cal V}^{2/3}}{\tau_\phi}\right)\right].
\end{equation}
Here $f$ is a generic function of the variable ${\cal V}^{2/3}/\tau_\phi$ 
and $\cloop$ is a numerical factor to be discussed in more detail below. 

Crucially, the $\tau_\phi$-independent term $|W_0|^2/{\cal V}^3 \times {\cal O}(1)$ in \eqref{eq:potential-prediction} arises because we assume that, at the end of inflation, the additional K\"ahler modulus $\tau_\phi$ settles in a stable minimum. For a blow-up cycle $\tau_\phi$, such a minimum is generically induced by non-perturbative effects, in analogy to the minimum for $\tau_s$. Alternatively, a stabilization by perturbative effects is also possible \cite{Cicoli:2008va, Cicoli:2011qg, Cicoli:2012sz}. The depth of this minimum is $\sim |W_0|^2/{\cal V}^3$, which is larger than the loop-induced potential that in the inflationary region scales as $\sim |W_0|^2/{\cal V}^{10/3}$. We are thus dealing with a relatively flat inflationary plateau. Its profile is determined by string loops and the relation between $\tau_\phi$ and the canonically normalized inflaton $\phi$. Together this suggests the name `Loop Blow-up Inflation', characterizing the main features of our model.

Recalling that the metric 
$\partial^2 K /(\partial \tau_I \partial \tau_J)$ on K\"ahler moduli space is homogeneous of degree $-2$ in the $\tau_I$, one easily shows that:
\begin{equation}
	\epsilon\equiv\frac{1}{2}\left(\frac{1}{V(\phi)}\frac{dV(\phi)}{d\phi}\right)^2\sim \left(\frac{c_\text{loop}}{{\cal V}^{1/3}} \, \frac{\dd f}{\dd \phi} \right)^2\,\,\,,\qquad 
    \eta \equiv \frac{1}{V(\phi)}\frac{\dd^2V(\phi)}{\dd\phi^2} \sim \frac{c_\text{loop}}{{\cal V}^{1/3}} \, \frac{\dd^2 f}{\dd \phi^2}\,.
    \label{srp}
\end{equation}
In the region in moduli space where $\tau_\phi\lesssim \mathcal{V}^{2/3}$, we have $f'' \sim f' \sim {\cal O}(1)$. Slow-roll inflation is hence parametrically guaranteed by ${\cal V}\gg 1$. 

Note that all of the above goes through if, on top of the blow-up mode $\tau_\phi$, there are $(n-1)$ additional K\"ahler moduli. The inflationary plateau is still sufficiently flat, being described by \eqref{eq:potential-prediction} where now $f$ is a generic function of the $n$ variables ${\cal V}^{2/3}/\tau_i$ by performing the replacement $f\left({\cal V}^{2/3}/\tau_\phi\right)\to
f\left({\cal V}^{2/3}/\tau_i\right)$. The crucial ${\cal O}(1)$ term in \eqref{eq:potential-prediction} arises when, after the end of inflation, all the $\tau_i$ eventually settle in a deep minimum. We emphasize that the key result of \eqref{srp} continues to hold, with $f$ that now has to be interpreted as the single-variable function obtained from $f(\vol^{2/3}/\tau_i)$ by restriction to the canonical field $\phi$ parameterizing the inflationary trajectory on the $n$-dimensional plateau. Of course, all moduli need to be stabilized which, while it will in general happen due to either loop or non-perturbative effects, does not allow us to be as explicit as in the case of a single field~$\tau_\phi$. Moreover, a multi-field analysis would require to study the evolution of isocurvature perturbations, which is beyond the scope of our paper.

We therefore focus on the case of just one blow-up modulus $\tau_\phi$ and demand that the overall volume remains approximately fixed during inflation, i.e.~that our inflationary model is effectively single-field. This can be achieved if the effects that stabilize $\tau_\phi$ are subdominant with respect to the leading order potential even when $\tau_\phi$ is close to the minimum. When $\tau_\phi$ is fixed by non-perturbative effects, these can be subdominant due to 3 reasons: ($i$) leading order instanton contributions could be absent due to too many fermionic zero modes or a prefactor proportional to vanishing matter field VEVs (as in the presence of gauge fluxes); ($ii$) the rank $N_\phi$ of the condensing gauge group relevant for $\tau_\phi$ could be much smaller than the rank $N_s$ relevant for $\tau_s$: $\,N_\phi\ll N_s$; ($iii$) the Calabi-Yau could feature $n_s\gg 1$ small-cycle moduli $\tau_s$, as in \cite{Conlon:2005jm}. On the other hand, when $\tau_\phi$ is fixed by string loops\footnote{As proposed in \cite{Cicoli:2008va, Cicoli:2011qg, Cicoli:2012sz}, $\tau_\phi$ might be fixed by string loops relative to $\tau_s$ which is stabilized non-perturbatively. \label{fln}}, the extended no-scale cancellation guarantees that these contributions are suppressed.

The rest of this paper is structured as follows. In Sec.~\ref{sc:simplest-setting} we study the simple special case with a single additional K\"ahler modulus, $\tau_i\equiv \tau_\phi$, which, as explained above, is of blow-up type. In this setting the form of the dominant loop correction in the regime $\tau_s\ll \tau_\phi\lesssim {\cal V}^{2/3}$ is actually known from an explicit analysis in \cite{Gao:2022uop}, consistently with the extrapolation from the toroidal orientifold case by the conjecture proposed in~\cite{Berg:2007wt}. Then, approaching the regime of $\tau_\phi \lesssim \vol^{2/3}$ from the side of small $\tau_\phi$, we may hope to maintain control of the inflationary potential while also achieving realistic phenomenology. This allows for a very explicit case study where we derive the inflationary predictions for the simplest realization of Loop Blow-up Inflation.
The setting may be viewed as deriving from the original model \cite{Conlon:2005jm} by taking the (naively fatal) loop corrections into account and saving inflation at the price of moving to much larger values of $\tau_\phi$. Sec. \ref{sc:control-constraints} deals with questions of control over the EFT after finding the values of the microscopic parameters which match CMB data. We continue in Sec. \ref{sec:analysing-different--inflationary-regimes} by first exploring more general possibilities for the functional form of the loop corrections, and then by quantifying how small loop corrections would have to become to make a transition to the original model of \cite{Conlon:2005jm} where the inflationary potential is generated by non-perturbative effects. Finally, a detailed phenomenological analysis, including reheating, dark radiation constraints and an estimate of the inflationary parameters is given in Sec.~\ref{sc:pheno}.

\section{Loop blow-up inflation}
\label{sc:simplest-setting}

\subsection{The simplest model}

Our goal is to implement the central idea outlined in Sec.~\ref{sc:intro} using a concrete and simple example. For this, we choose the volume to have the form:
\begin{equation}
    \label{eq:Swiss-volume}
    \vol \,\,=\,\, 
    \tilde{\vol}(\tau_b,\tau_\phi)-\lambda_s\tau_s^{3/2} \,\,=\,\,
    \tau_b^{3/2}  -\lambda_\phi \tau_\phi^{3/2}
    - \lambda_s \tau_s^{3/2}~.
\end{equation}
In other words, we assume that in addition to the big and small cycles $\tau_b$ and $\tau_s$ of the LVS~\cite{Balasubramanian:2005zx,Conlon:2005ki}, there is just one further 4-cycle $\tau_\phi$ and that the latter is of blow-up type.

Let us discuss our setup and notation in more detail. In the above, $\tau_i$ are the real parts of the \Kahler moduli
\begin{equation}
    T_i = \tau_i + i ~c_i~,\quad i\in\{b,\,s,\,\phi\}\,,
\end{equation}
with $c_i$ their axionic partners and the constants $\lambda_s$ and $\lambda_\phi$ represent ratios of triple intersection numbers.
The \Kahler potential $K$, including the leading $\alpha'^3$ correction~\cite{Balasubramanian:2005zx,Becker:2002nn}, reads
\begin{equation}
\label{eq:kahler-potential}
    K =  K_{\rm cs}  - 2 \ln({\vol} +\hat{\xi}/2)\,,
\end{equation}
with $K_{\rm cs}$ depending only on complex structure moduli and axio-dilaton. Since these are stabilized by fluxes~\cite{Giddings:2001yu}, $K_{\rm cs}$ can be treated as a constant.
Furthermore, we have \mbox{$\hat{\xi}=-\zeta(3)\,\chi\big/[2\,(2\pi)^3\,g_s^{3/2}]$}, where $\chi$ denotes the Calabi-Yau Euler number. Note that $N=1$ corrections in general induce moduli redefinitions at $\mathcal{O}(\alpha'^2)$ \cite{Grimm:2013gma} and a shift in $\chi$ at $\mathcal{O}(\alpha'^3)$~\cite{Minasian:2015bxa}. The superpotential is given by
\begin{equation}
    \label{eq:superpotential}
    W = W_0 +  A_s \,e^{-a_s T_s} +A_\phi\, e^{-a_\phi T_\phi}~,
\end{equation}
where the constant contribution $W_0$ is generated by fluxes and the non-perturbative corrections arise from E3-branes ($a_{s,\,\phi}=2\pi$) or through gaugino condensation ($a_{s,\,\phi}=2\pi/N_\phi$). The prefactors $A_{s}$ and $A_\phi$ are ${\cal O}(1)$ numbers which depend on the complex structure moduli.

The super- and K\"ahler potential give rise to the $F$-term scalar potential for the K\"ahler moduli
\begin{gather}
\label{eq:potential}
    V(\vol, \tau_s,\tau_\phi) = V_\text{LVS}({\vol},\tau_s) + \hat{V}  \left[~ \A_\phi\, \frac{\sqrt{\tau_\phi} ~e^{-2 a_\phi \tau_\phi}}{{\vol}} - \B_\phi \, \frac{\tau_\phi ~ e^{-a_\phi\tau_\phi}}{{\vol}^2} \right]~,
\end{gather}
where $V_{\rm LVS}$ is the scalar potential of the underlying 2-moduli LVS model
\begin{equation}
    \label{eq:LVS-potential}
    V_\text{LVS}({\vol},\tau_s) = \hat{V}  \left[~ \A_s\, \frac{\sqrt{\tau_s} ~e^{-2 a_s \tau_s}}{{\vol}} - \B_s \, \frac{\tau_s ~ e^{-a_s\tau_s}}{{\vol}^2} +  \frac{3 \hat{\xi}}{4{\vol}^3}   \right]\,,
\end{equation}
and
\begin{equation}
    \label{eq:LVS-potential-prefactors}
    \hat{V} \equiv \left( \frac{\gs e^{K_{cs}}}{8\pi} \right)  W_0^2\,,\qquad
    \A_i \equiv \frac{8 (a_i A_i)^2}{3{W_0^2}\lambda_i}\,,\qquad
    \B_i \equiv 4 \frac{a_i |A_i|}{W_0}~,
\end{equation}
with $i = {s,\phi}$ labelling the blow-up cycles. Famously, the potential \eqref{eq:LVS-potential} has an AdS minimum at $\tau_s\sim (\hat{\xi}/2\lambda_s)^{2/3}$ and ${\cal V}\sim \exp(a_s\tau_s)$. In the full potential \eqref{eq:potential}, the additional $\tau_\phi$-dependent terms stabilize $\tau_\phi$ such that $a_\phi\tau_\phi\sim \ln {\cal V}$, analogously to $\tau_s$. Moreover, if we assume \cite{Conlon:2005jm, Cicoli:2016olq}
\begin{equation}
    \label{eq:displacement-condition}
    \lambda_\phi a_\phi^{-3/2}\ll
    \lambda_s a_s^{-3/2}\,,
\end{equation}
then the presence of $\tau_\phi$ and its stabilization do not affect the values of ${\cal V}$ and $\tau_s$ derived from $\eqref{eq:LVS-potential}$. This remains true even during inflation, when $\tau_\phi$ is displaced from its late-time AdS minimum. The AdS minimum may be uplifted to a Minkowski minimum by adding to the potential in \eqref{eq:potential} a positive term which can be parametrized as:
 \begin{equation}
 \label{eq:uplift-def}
     V_\text{up}(\vol) = \frac{\hat{V}\D}{\vol^2}\,.
 \end{equation}
This term is such that $(V+V_\text{up})\big|_\text{minimum}=0$.\footnote{
For 
a precise determination of the constant $\D$ see eq.~(4.7) of \cite{Cicoli:2016olq}.} We note that, while the feasibility of the famous anti-D3-brane uplift \cite{Kachru:2002gs,Kachru:2003aw} has been challenged in this context \cite{Junghans:2022exo,Gao:2022fdi,Junghans:2022kxg,Hebecker:2022zme,Schreyer:2022len,Schreyer:2024pml}, we are here simply assuming that {\it some} form of viable uplift for the LVS can be realized, as in~\cite{Saltman:2004sn,Cremades:2007ig,Cicoli:2012fh,Braun:2015pza,Retolaza:2015nvh,Cicoli:2015ylx,Gallego:2017dvd,Hebecker:2020ejb,Krippendorf:2023idy} which proposed alternative uplifting mechanisms.
 
Note that we have arranged the expression for the potential in \eqref{eq:potential} so that one can clearly distinguish the standard LVS scalar potential $V_\text{LVS}$, independent of the additional modulus $\tau_\phi$, and the non-perturbative corrections giving $\tau_\phi$ a non-trivial potential.
If no further terms were added, $\tau_\phi$ could be the inflaton of the original model \cite{Conlon:2005jm}.
In this case, an inflationary plateau appears in the region where $\tau_\phi$ is large enough for the exponential terms to become sufficiently small. Our proposal is different: we will include loop corrections, making the potential for $\tau_\phi$ less flat but, in the regime where $\tau_\phi$ comes close to ${\cal V}^{2/3}$, still suitable for slow-roll inflation. In fact, we will argue that this is an unavoidable outcome. In other words, `blow-up inflation'~\cite{Conlon:2005jm} {\it necessarily} turns into a variant of what we would like to call `Loop Blow-up Inflation'.

Let us be more explicit by specifying the leading loop correction to the potential, as it arises from a loop effect in the \Kahler potential $K$:
\begin{equation}\label{eq:loop-corrections}
    \delta V_\textrm{loop} \simeq -
        \frac{\hat{V}}{\vol^3}~\frac{c_\text{loop}}{ \vol^{1/3}} ~f\left(\frac{\vol^{2/3}}{\tau_\phi}\right). 
\end{equation}
Here $f$ is a generic function of ${\cal V}^{2/3}/\tau_\phi$. The full potential $V$ hence reads:
\begin{equation}
    \label{eq:full-potential}
    V(\vol, \tau_s,\tau_\phi) = V_\text{LVS}({\vol},\tau_s) +V_\text{up}(\vol) + \hat{V} \left[~ \A_\phi\, \frac{\sqrt{\tau_\phi} ~e^{-2 a_\phi \tau_\phi}}{{\vol}} - \B_\phi \, \frac{\tau_\phi ~ e^{-a_\phi\tau_\phi}}{{\vol}^2} \right] + \delta V_\textrm{loop}~.
\end{equation}
Two key points have to be made concerning this potential: one concerning our claim that a $\tau_\phi$-dependent correction $\delta V_\textrm{loop}$ is unavoidable, and another, closely related point concerning the form of this correction as well as the form of the function $f$ in \eqref{eq:loop-corrections}.

We start with the claim that such a correction is unavoidable. Indeed, to realize the minimum which stabilizes $\tau_\phi$ after inflation, we require that $W$ receives a non-perturbative correction $\sim \exp(-a_\phi T_\phi)$, cf.~\eqref{eq:superpotential}.\footnote{If the non-perturbative corrections to $W$ are very suppressed (as in the case where leading order instantons vanish), another possibility to generate the minimum for $\tau_\phi$ would be to use additional loop corrections -- cf.~footnote \ref{fln} in the Introduction.}  Requiring this non-perturbative correction implies the presence of an O-plane in the vicinity of the blow-up cycle $\tau_\phi$ in order to break SUSY locally to ${\cal N}=1$.\footnote{
Note that fluxes can break SUSY to ${\cal N}=1$ as well. However, this does not introduce non-perturbative corrections to $W$. The reason is that fluxes become diluted as the volume ${\cal V}=\Re T_b$ grows. Hence the corrections would have to be of the form $\A({\cal V})\exp(-a_\phi\tau_\phi)$. This is ruled out by the holomorphy of $W$ in $T_b$. We thank T. Weigand for pointing this out. A related argument can be found in Sec.~3.2 of \cite{Conlon:2005jm}.

It is conceivable that, even in the absence of a local O-plane, a non-perturbative minimum stabilizing $\tau_b$ arises due to Beasely-Witten $F$-terms \cite{Beasley:2005iu, Blumenhagen:2009qh}. We expect that in this case both the height of the inflationary plateau and the size of loop corrections on the plateau are reduced. Whether this can lead to a realistic model of inflation remains to be seen.
} 
As has been discussed in detail in \cite{Gao:2022uop}, this locally reduced SUSY then also implies the presence of the claimed loop effect.

Since this last point is crucial, we want to provide more details. Recall first that it has been argued in EFT language that corrections suppressed by ${\cal V}^{10/3}$ arise from 10d field-theory loops in ${\cal N}=1$ CY orientifold models [\citealp{vonGersdorff:2005bf},~\citealp{Cicoli:2008va}]. At the same time, this has been derived in a very impressive, explicit string-loop calculation, which is however necessarily restricted to torus-based geometries \cite{Berg:2005ja}. A generalization to the CY case was conjectured in~\cite{Berg:2007wt}, and \cite{Cicoli:2007xp} provided a low-energy interpretation of one-loop open string corrections by matching them with the one-loop Coleman-Weinberg potential. This was developed and partially debated in \cite{Gao:2022uop}. We provide more details on this and on the effect of fluxes on the loop corrections in App.~\ref{app:loop-notation}. Not to lose focus, we state here only that, even in the absence of open string loops as in the case when the relevant cycle is not wrapped by any D7-brane, one-loop closed string effects unavoidably induce a correction of the type given in \eqref{eq:loop-corrections} as soon as the relevant geometry breaks SUSY to ${\cal N}=1$. 

Moreover, the precise functional form of $f$ in \eqref{eq:loop-corrections} in an explicit Calabi-Yau setting is unknown. Yet, in the regime where a blow-up cycle $\tau_\phi$ is smaller than any other nearby cycle, one can argue in effective field theory for a loop correction depending on $\tau_\phi$ only and, through Weyl rescaling of the 4d metric, on ${\cal V}$. As estimated in \cite{Cicoli:2007xp} for open string loops and as derived in \cite{Gao:2022uop} for closed string loops, this leads to:
\begin{equation}
\label{eq:leading-loop-correction}
    f\simeq \frac{\vol^{1/3}}{\sqrt{\tau_\phi}}\qquad \quad \mbox{and hence} \qquad \quad
    \delta V_\textrm{loop} \simeq -\frac{\hat{V}}{\vol^3}~\frac{c_\text{loop}}{ \sqrt{\tau_\phi}}\,.
\end{equation}
Here any unknown ${\cal O}(1)$ factors in $f$ have been absorbed in $c_\textrm{loop}$.
We also note that this numerical factor, which does not involve $g_s$, is expected to be small in (higher-dimensional) analogy to the familiar loop suppression factor $1/(16\pi^2)$ of 4d field theory. Reference~\cite{Gao:2022uop} derives the value $1/(2\pi)^4$ from the explicit torus orbifold results of \cite{Berg:2005ja}. Alternatively, an identification of the relevant cutoff with the Kaluza-Klein scale naively given by $M_p/(\tau_\phi^{1/4}\sqrt{\cal V})$, allows one to use the 4d value $1/(16\pi^2)$. We will use the latter, more conservative value.

As explained above in relation to \eqref{eq:displacement-condition}, we may choose CY data such that our potential inflaton $\tau_\phi$ can roll while 
$\vol$ and $\tau_s$ remain stabilized (up to small shifts) \cite{Conlon:2005jm,Cicoli:2016olq}.
We may then work with a potential depending on $\tau_\phi$ only:
\begin{equation}
\label{eq:tauphi-pot}
    V(\tau_\phi) = V_0 \left[ 1 + \A_\phi\frac{{\vol^2}}{\beta} ~\sqrt{\tau_\phi} ~e^{-2 a_\phi \tau_\phi} -  \B_\phi\frac{{\vol}}{\beta} ~ \tau_\phi ~ e^{-a_\phi\tau_\phi} - \frac{c_\text{loop}}{\beta\sqrt{\tau_\phi}}\right]\,.
\end{equation}
Here we defined
\begin{equation}
        \label{eq:V0}
    V_0 \equiv \left[V_\text{LVS}(\vol, \tau_s) + V_\text{up}(\vol)\right] \Big|_\text{minimum} = \frac{\hat{V} \beta}{\vol^3}\,,
\end{equation}
with $\beta$ given by \cite{Cicoli:2016olq}:
\begin{equation}
\label{eq:beta-def}
    \beta \simeq \frac32 a_\phi^{-3/2} \lambda_\phi \left(\ln \vol\right)^{3/2}\,.
\end{equation}
The constant $\beta$ encodes the proper adjustment of the uplifting term \eqref{eq:uplift-def}, ensuring that $V_0$ precisely compensates the negative value arising from the two exponential terms in \eqref{eq:tauphi-pot} after minimization in $\tau_\phi$. Obviously, the resulting value of $\beta$ is corrected due to the presence of the $c_{\rm loop}$ term, but this is not important at our level of precision.

\begin{figure}[ht]
    \centering
    \includegraphics[width=0.95 \textwidth]{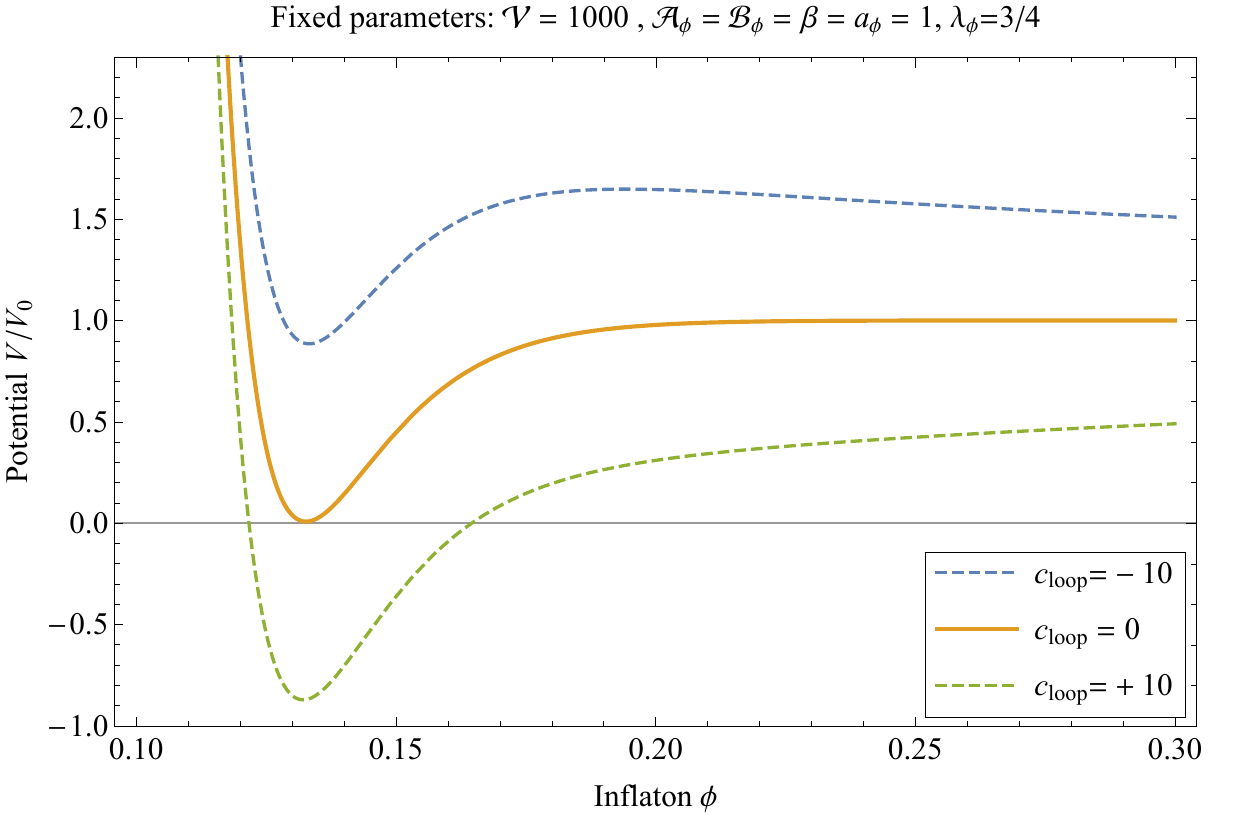}
    \caption{Plot of the potential \eqref{eq:tauphi-pot} for $\cloop=\pm10$ and $\cloop=0$.}
    \label{fig:potential}
\end{figure}

We identify the inflaton $\phi$ with the canonically normalized field corresponding to $\tau_\phi$:
\begin{equation}
\label{eq:canonically normalized inflaton}
    \phi = \sqrt{\frac{4 \lambda_\phi}{3{\vol}}} ~\tau_\phi^{3/4}\,.
\end{equation}
In terms of $\phi$, the full inflationary potential~\eqref{eq:tauphi-pot} has several regimes which allow to realize slow-roll inflation. To begin this discussion, we present in Fig.~\ref{fig:potential} a plot of our potential~\eqref{eq:tauphi-pot} for different values of $\cloop$. The orange curve corresponds to $c_{\rm loop}=0$ and is adjusted such that the minimum is at zero energy. The blue and green curves have positive and negative $c_{\rm loop}$ respectively. Obviously, when applying either of them to cosmology, the constant term must be adjusted such that its minimum (rather than that of the orange curve) is Minkowski. Note that we used extreme values for $\cloop$ in Fig.~\ref{fig:potential} to make the loop effect more visible. 

As is well-known and also visible in the plot, the pure blow-up case with $c_{\rm loop}=0$ has a slow-roll region which starts relatively close to the minimum. The reason is that the potential approaches a constant exponentially fast. As we argued above and will quantify later in Sec.~\ref{sc:Original-blow-up-regime}, the loop correction destroys this slow-roll region if $c_{\rm loop}\gtrsim 10^{-6}$. In case $c_{\rm loop}>0$, slow-roll can then be regained at much larger $\phi$.  For $c_{\rm loop}<0$, this is impossible.

For the case $\cloop\gtrsim 10^{-6}$, if $\tau_\phi$ is displaced within the regime $\tau_s\ll\tau_\phi<\tau_b$ so that the two exponential terms in \eqref{eq:tauphi-pot} can be neglected, 
the potential in terms of $\phi$ takes the form:
\begin{equation}
    \label{eq:fullpotential}
    V(\phi) = V_0 \left(1 - \,\frac{b\,c_\text{loop}}{\phi^{2/3}} \right)\qquad\text{where}\qquad b\equiv \frac{1}{\beta}\,\left(\frac{4 \lambda_\phi}{3\vol}\right)^{1/3}\equiv\frac{\sigma_\phi}{\beta\vol^{1/3}}\,.
\end{equation}
This characterizes the slow-roll regime in our simplest scenario which relies on loop corrections to drive inflation.

In the rest of this section and Sec.~\ref{sc:control-constraints},
we assume that the approximate potential of~\eqref{eq:fullpotential} can be used in the inflationary regime. This implies in particular that $\tau_\phi$ is small enough such that the leading-order term in the expansion of $f$ in $\tau_\phi$ in (\ref{eq:leading-loop-correction}) is sufficient. Yet, we want to emphasize that this is merely one regime in which slow-roll can be realized. In Sec.~\ref{sec:analysing-different--inflationary-regimes}, we study two additional slow-roll regimes: the regime where subleading terms in the small-$\tau_\phi$ expansion of $f$ are relevant, and the regime where loop corrections become negligible due to a small value of $\cloop$.

\subsection{Inflationary dynamics}
\label{sc:inflation-parameters-simplest-setting}

In this subsection we assume that the approximate potential \eqref{eq:fullpotential} is sufficient to describe the observable part of slow-roll inflation. This implies that the exponentially suppressed terms are negligible, which will always hold as long as $c_{\rm loop}$ is not too small. It also relies on the requirement:
\begin{equation}
\tau_\phi \lesssim {\vol}^{2/3}
\qquad \quad \mbox{or equivalently}
\quad \qquad
\phi \lesssim 1\,,
\label{eq:upper bound on phi}
\end{equation}
in order to realize inflation within the K\"ahler cone. The requirement that $\tau_\phi$ is far away from the walls of the K\"ahler cone during the observable $\sim 52$ efoldings of inflation constrains $c_{\rm loop}$ from above.

The slow-roll parameters following from the potential \eqref{eq:fullpotential} read:
\begin{align}
    & \epsilon = \frac{1}{2}\left(\frac{V_\phi}{V}\right)^2 \simeq \frac{2}{9}\,\frac{(b\,c_\text{loop})^2}{\phi^{10/3}}\,, \label{eq:epsilon corrected}\\
    & \eta  = \frac{V_{\phi\phi}}{V} \simeq -\frac{10}{9}\,\frac{b\,c_\text{loop}}{\phi^{8/3}}\,. \label{eq:eta corrected}
\end{align}
We can see that for small values of $(b\,\cloop)$ a slow-roll regime can be realized.
With the parameters $\epsilon$ and $\eta$ at hand, we can determine the spectral index $n_s$, the tensor-to-scalar ratio $r$ and the number of efoldings $N_e$: 
\begin{align}
    n_s &= 1+ 2 \,\eta - 6\, \epsilon \simeq 1 -\frac{20}{9}\, \frac{b\, \cloop}{\phi_*^{8/3}}\,,\label{eq:n_s corrected}\\
    r& = 16\,\epsilon \simeq \frac{32}{9}\, \frac{(b\, \cloop)^2}{ \phi_*^{10/3}}\,, 
    \label{r} \\
    \label{eq:e-foldings}
    N_e &= \int_{\phi_{\rm end}}^{\phi_*} \frac{V}{V_\phi} \, \dd \phi \simeq \frac{9}{16} \frac{\phi_*^{8/3}}{b\,\cloop}\,,
\end{align}
where $\phi_{\rm end}$ and $\phi_*$ denote respectively the values of the inflaton field at the end of inflation (where $\epsilon\simeq 1$) and at the scale of horizon exit. In \eqref{eq:e-foldings} we used $\phi_{\rm end}\ll\phi_*$. 

Based on the formulae above, we will aim to match cosmological constraints in the next section. Specifically, we need to ensure the right number of efoldings $N_e$ and the right amplitude of primordial density fluctuations $\tilde{A}_s$.
These requirements will fix $\phi_*$ and $\V$ in terms of $N_e$ and $\tilde{A}_s$.
The hope is now that the large parameters $N_e$ and the inverse spectrum normalization $\tilde{A}_s^{-1}$ are sufficient to make the volume $\V$ large enough to realize a controlled LVS model. At the same time, the condition $\phi_*\lesssim 1$ has to be maintained.

\section{Control and constraints}
\label{sc:control-constraints}

In this section we implement phenomenological constraints on the microscopic parameters of our loop blow-up inflation model, focusing on its simplest realization.
The inflationary parameters specific to this scenario have been obtained in Sec.~\ref{sc:inflation-parameters-simplest-setting}. Our primary objective is to assess whether the modulus $\tau_\phi$  corresponding to the inflaton remains inside the K\"ahler cone, which is equivalent to $\phi\lesssim 1$. In Sec.~\ref{sc:para-control}, we derive formulas for $\phi_*$ and $\V$ in terms of the number of efoldings $N_e$ and the amplitude of the density perturbations $\hat{A}_s$. Following this, Sec.~\ref{KaehlerCone} will show, using the number of efoldings, $N_e\simeq 52$, and typical values of the volume, $\vol\sim\mathcal{O}(10^4)$, derived in Sec. \ref{sc:pheno}, that our model indeed remains in the controlled regime. Moreover, in Sec.~\ref{sc:warping-corr} we argue that achieving a Minkowski vacuum via anti-D3-brane uplift is challenging due to 10d curvature corrections. 

\subsection{Matching cosmological data}
\label{sc:para-control}

First, we will discuss how $\vol$ and $\phi_*$ are constrained by cosmological observables. Our first constraint arises from matching the amplitude of primordial density fluctuations $\tilde{A}_s$. The spectrum of scalar density perturbations $\Delta_s^2$ is defined by (cf.~eq.~(1.58) in \cite{Baumann:2014nda})
\begin{equation}
    \Delta_s^2 = \tilde{A}_s \left( \frac{k}{k_*}\right)^{n_s-1}\,,
\end{equation}
where the amplitude $\tilde{A}_s$ was measured by Planck \cite{Planck:2018vyg}:
\begin{equation}
    \tilde{A}_s\times10^{9} = 2.105 \pm 0.030\,.
\end{equation}
We also have
(cf. eq. (2.42) in \cite{Baumann:2014nda})
\begin{equation}    \label{eq:scalarnormalisation}
    \Delta_s^2 =\left. \frac{1}{24 \pi^2} \frac{V}{\epsilon}\right\rvert_{\phi=\phi(k)}~,
\end{equation}
where $V$ is the scalar potential. Evaluating \eqref{eq:scalarnormalisation} at $ \phi(k_*)=\phi_* $ is thus equivalent to:
\begin{equation}
\label{eq:scalar-perturbations-potential}
    \left. \frac{V^3}{V_\phi^2}\right\rvert_{\phi=\phi_*} = \hat{A}_s \equiv 12\pi^2 \tilde{A}_s \simeq 2.5 \times 10^{-7} \,.
\end{equation}
Using the potential \eqref{eq:fullpotential} and the approximation $1-\cloop b \phi_*^{-2/3}\simeq 1$ in \eqref{eq:scalar-perturbations-potential} yields:
\begin{equation}
\label{eq:scalar-perturbation-matching}
    \frac{9V_0}{4} \frac{\phi_*^{10/3}}{(b\,\cloop)^2} = \hat{A}_s\,.
\end{equation}
Recall that $V_0$ and $b$ contain the volume $\V$, cf. \eqref{eq:V0} and \eqref{eq:fullpotential}. Thus, we interpret \eqref{eq:scalar-perturbation-matching} as a relation between the value of the inflaton at horizon exit $\phi_*$  and the volume $\V$.

In a second step, the required number of efoldings $N_e$ is determined by the post-inflationary history specific to the inflation model. We will perform this analysis for our particular model in Sec.~\ref{sc:pheno}, where we will find $N_e \simeq 51.5$-$53$ depending on the underlying brane setup. Thus, we can essentially treat $N_e$ as a fixed constant and interpret  \eqref{eq:e-foldings} as a second constraint on $\phi_*$ and ${\cal V}$. The two relations \eqref{eq:e-foldings} and \eqref{eq:scalar-perturbation-matching} can be solved for $\phi_*$ and $\vol$: First, we solve \eqref{eq:e-foldings} for $\V$ in terms of $\phi_*$:
\begin{equation}
\label{eq:volume-sol-simple}
\V=\frac{A}{\phi^{8}_*}\,,\qquad A \equiv \left( \frac{16 N_e \sigma_\phi \cloop}{9 \beta} \right)^3\,.
\end{equation}
Next, inserting this expression for $\V$ in \eqref{eq:scalar-perturbation-matching} we find:
\begin{equation}
\label{eq:phi-sol-simple}
 \phi_*= \left( B A^7 \right)^{\frac{1}{66}} ~,\qquad B \equiv \left( \frac{4 \hat{A}_s \sigma_\phi^2 \cloop^2}{9 \beta^3 \hat{V}} \right)^3~.
\end{equation}
Thus, $\phi_{*}$ and $\V$ are expressed in terms of $A$ and $B$.
Using the definitions of $A$ and $B$ gives
\begin{gather}
\label{eq:solutions-NQ}
    \phi_*= \left(\frac{2^{17} \pi}{3^{8}}\right)^{\frac{1}{11}} \left[   \frac{\hat{A}_s \, N_e^{7}\left(\sigma_\phi\,\cloop\right)^9}{N_Q~\beta^{10}}\right]^{\frac{1}{22}},\qquad 
\vol=  \left[ \frac{1}{144\,\pi^8} \frac{N_e^{5}\, N_Q^4 \,\beta^{7}}{\hat{A}_s^4 \left(\sigma_\phi\, \cloop\right)^3} \right]^{\frac{1}{11}}  \,,
\end{gather}
where we introduced the convenient parameter
\begin{gather}
\label{eq:flux-tadpole-contribution}
N_Q\equiv 2\pi\gs e^{K_{cs}}W_0^2\,,
\end{gather}
which contains all $W_0$ and $\gs$ dependencies.
Note that the quantity $N_Q$ is bounded by the negative tadpole $-Q_3$ of the orientifold~\cite{Denef:2004ze}:
\begin{equation}
\label{eq:DD-bound}
    N_Q<-Q_3\sim {\cal O}(100)\,.
\end{equation}
Hence our ability to raise $\vol$ and lower $\phi_*$ by using large $N_Q$ is limited. In type IIb models with local D7-tadpole cancellation, an upper bound on $-Q_3$ follows from the Lefschetz fixed point theorem \cite{Collinucci:2008pf,Carta:2020ohw,Bena:2020xrh}, relying solely on the Hodge numbers of the CY. The largest numbers from the Kreuzer-Skarke database \cite{Kreuzer:2000xy} imply a maximum at $-Q_3=252$, which was also explicitly realized in orientifold models in~\cite{Crino:2022zjk}.

Parametrically, a small value for $\phi_*$ and a large volume $\vol$ can be achieved due to the small amplitude $\hat{A}_s$ and the small factor $\cloop$. In addition, $N_Q\sim \O(100)$ can provide further but limited improvement. For illustrative purposes, let us adopt the following natural choice of microscopic parameters:\footnote{Imposing perturbative control by requiring $g_s \lesssim 0.2$ implies, for our parameter choice, $e^{K_{\rm cs}}\gtrsim 10$.}
\begin{eqnarray}
\lambda_\phi=1\,,\qquad c_{\rm loop}= 1/(16\pi^2)\,,\qquad \beta=W_0=g_s\,e^{K_{\rm cs}}=2 \quad \Rightarrow\quad N_Q=16\pi\,.
\label{Parameters}
\end{eqnarray}
For this parameter choice and $N_e \simeq 51.5$-$53$, (\ref{eq:solutions-NQ}) reduces to:
\begin{equation}
\phi_* = 0.06\, N_e^{7/22}\sim \mathcal{O}(0.2) \qquad\text{and}\qquad 
\vol =  1743\, N_e^{5/11}\sim \mathcal{O}(10^4)\,.
\label{VolDetermination}
\end{equation}
Thus, one may say that Loop Blow-up Inflation is typically characterised by $\phi_*\sim \mathcal{O}(0.2)$ and $\vol\sim \mathcal{O}(10^4)$. Compared with the original model based on non-perturbative corrections~\cite{Conlon:2005jm}, $\phi_*$ is much larger while $\vol$ is slightly smaller.

\subsection{K\"ahler cone constraints}
\label{KaehlerCone}

We now want to estimate whether we can realize slow-roll, remain inside the K\"ahler cone and simultaneously match the constraints discussed in Sec.~\ref{sc:para-control}. As an illustrative example and without loss of generality, we take the explicit construction of \cite{Cicoli:2012vw} which features a CY volume of the form (setting one exceptional divisor to zero size)
\begin{equation}
    \vol = \frac19\sqrt{\frac23}\left(\tau_b^{3/2}- \sqrt{3}\,\tau_s^{3/2}-\sqrt{3}\,\tau_\phi^{3/2}\right),
\end{equation}
with the following relations between 4-cycle and 2-cycle volumes
\begin{equation}
\tau_b = \frac{27}{2}\,t_b^2\,,\qquad \tau_s = \frac92\,t_s^2\qquad \tau_\phi=\frac92\,t_\phi^2\,,
\end{equation}
and K\"ahler cone conditions
\begin{equation}
t_b+t_s>0\,,\qquad t_b+t_\phi>0\,,\qquad t_s<0\,,\qquad t_\phi<0\,.
\end{equation}
The canonical normalization \eqref{eq:canonically normalized inflaton} therefore becomes
\begin{equation}
\label{eq:tau_phi*in terms of phi*}
    \tau_\phi = \left(\frac{\sqrt{3}}{4}\right)^{2/3} \V^{2/3} \phi^{4/3} \simeq \left(\frac{1}{18\sqrt{2}}\right)^{2/3}\tau_b\, \phi^{4/3} \,,
\end{equation}
where in the second equality we have approximated $\V \simeq \frac19\sqrt{\frac23}\,\tau_b^{3/2}$. We can then re\-write~\eqref{eq:tau_phi*in terms of phi*} in terms of the 2-cycles as
\begin{equation}
\label{eq:relation between t_phi and t_B}
    \frac{|t_\phi|}{t_b} =  \left(\frac{1}{2\sqrt{6}}\right)^{1/3} \,\phi^{2/3} \simeq 0.6\,\phi^{2/3}\,.
\end{equation}
Evaluating this ratio at horizon exit we find
\begin{equation}
\label{eq:ratio of t_phi and t_B}
    \frac{|t_{\phi_*}|}{t_b} \simeq 0.6\,\phi_*^{2/3}\simeq 0.2\qquad\text{for}\qquad \phi_*\simeq 0.2\,,
\end{equation}
which implies that the whole inflationary dynamics takes place well inside the K\"ahler cone with $|t_\phi| \ll t_b$. For $\vol\sim \mathcal{O}(10^4)$, one has $|t_{\phi_*}|\sim \mathcal{O}(2.5)< t_b\sim \mathcal{O}(13)$.

\subsection{10d curvature corrections}
\label{sc:warping-corr}

Furthermore, we face another consistency constraint on the volume $\V$ if the uplift mechanism relies on warped throats.
In the case of anti-D3-uplift, warped throats are present and we have to take a correction into account which arises as a combination of the leading $\ap^3$-correction and a non-constant warp factor \cite{Junghans:2022exo,Gao:2022fdi}.
Control over these corrections can be assured if the parameter $c_N$, as defined in (3.9) of \cite{Gao:2022fdi}, satisfies $c_N\gg1$. The parameter $c_N$ is given by
\begin{equation}
    \label{eq:cN-eq}
    c_N= \frac{\vol^{2/3}}{N} \frac{(2\, \lambda_s)^{2/3}}{10\, a_s\, \hat{\xi}^{2/3}}~,
\end{equation}
where $N$ denotes the D3-tadpole contribution from the fluxes in the throat.\footnote{The relation \eqref{eq:cN-eq} was derived using the volume in the global LVS minimum. However, as we already discussed in Sec.~\ref{sc:intro}, this is sufficiently similar to the volume $\V$ during inflation.}
We can rewrite \eqref{eq:cN-eq} purely in terms of the volume $\V$
\begin{equation}
    c_N= \frac{\vol^{2/3}}{N} \frac{1}{10}~\frac{1}{\ln \vol}\,,
\end{equation}
since $\tau_s$ stabilizes $\vol$ at $\vol \sim \exp(a_s \tau_s)$ where in addition $\tau_s\sim \hat{\xi}^{2/3}/(2\lambda_s)^{2/3} $.
Solving the above equation approximately by neglecting the $\ln\vol$ contribution yields:
\begin{equation}
    \vol = (10 N c_N)^{3/2}\,.
\end{equation}
The requirement of control, i.e. $c_N \gg1$, hence constrains $\V$ to satisfy
\begin{equation}
\label{eq:volumeBound}
    \vol \gg  (10\, N)^{3/2}\,.
\end{equation}
The minimal value for $N$ which allows for an anti-D3-brane uplift can be obtained from the parametric tadpole constraint (PTC) \cite{Gao:2022fdi}.
To obtain $N_\text{min}$ we use the minimal value for \mbox{$g_s M^2=144$} from \cite{Hebecker:2022zme,Schreyer:2022len} and apply the PTC \cite{Gao:2022fdi} which gives $N_\text{min}\approx 240$ such that we find the following lower bound on $\V$:
\begin{equation}
    \label{eq:volumeBound-phenoparamter}
    \vol \gg  10^5\left( \frac{N}{240}\right)^{3/2}~.
\end{equation}
 
Note that in the derivation of \eqref{eq:volumeBound} we neglected the logarithmic volume contribution and in the determination of $N_\text{min}$ we further omitted the subleading terms of the PTC.
Including any of these corrections will substantially worsen the bound \eqref{eq:volumeBound-phenoparamter} on $\V$ for our application. Therefore, if we insist on using anti-D3-brane uplift, our inflation model may run into trouble as our volume will not be large enough, cf. \eqref{VolDetermination}.
However, this constraint does not apply to alternative uplift mechanisms, like D-term effects \cite{Braun:2015pza}, dilaton-dependent non-perturbative contributions \cite{Cicoli:2012fh, Retolaza:2015nvh}, T-branes \cite{Cicoli:2015ylx} or non-zero F-terms of the complex structure moduli \cite{Gallego:2017dvd,Hebecker:2020ejb,Krippendorf:2023idy}. In addition, this constraint might not apply to another scenario which we will discuss in Sec.~\ref{sc:more-models-ratios} where we will go to the regime $\phi\sim\O(1)$ including subleading corrections.

\section{Further inflationary regimes}
\label{sec:analysing-different--inflationary-regimes}

The specific expression \eqref{eq:fullpotential} represents only one of the possible regimes in which the general potential \eqref{eq:loop-corrections} (with the loop correction defined in \eqref{eq:full-potential}) can realize slow-roll inflation. So far,
we assumed that this loop correction dominates over the exponential terms and thus represents the leading effect breaking the flatness of the potential. At the same time, we have remained in the small blow-up regime to maintain control over the explicit form of the function $f$, as specified in \eqref{eq:leading-loop-correction}. There are two alternative regimes, which arise as follows:

First, we may leave the small blow-up regime by going to $\phi\sim \mathcal{O}(1)$. In doing so we lose control over the explicit functional form of the loop correction.
However, as we approach this regime from small values of $\phi$, we can expand $f$ around the point $\tau_\phi/{\cal V}^{2/3}=0$, introducing subleading corrections and incorporating them into our analysis.
In Sec.~\ref{sc:more-models-ratios} we will argue that a whole class of inflationary models might arise in this regime. Second, it is possible that the loop factor $\cloop$ is so small that we can disregard loop corrections completely. This would bring us back to the original proposal of \cite{Conlon:2005jm}.
In Sec.~\ref{sc:Original-blow-up-regime} we will determine the critical value of $\cloop$ where this transition occurs.

\subsection{Subleading corrections}
\label{sc:more-models-ratios}

In this section, we consider the first alternative regime of the general potential \eqref{eq:full-potential}, which arises by moving to larger field values, $\phi\sim \mathcal{O}(1)$. This implies a departure from the specific functional form \eqref{eq:leading-loop-correction} of the leading loop correction. We assume that the leading loop correction can be interpreted as the first term of an expansion of $f$ in \eqref{eq:loop-corrections} in terms of the 2-cycle volume $\sqrt{\tau_\phi}$. Symbolically, the function $f$ then takes the form:
\begin{equation}
    \label{eq:corrections-in-f}
    f \simeq \frac{\vol^{1/3}}{\sqrt{\tau_\phi}}\left(1 + \frac{\sqrt{\tau_\phi}}{\vol^{1/3}} + \frac{\tau_\phi}{\vol^{2/3}}+\dots \right).
\end{equation}
The additional terms in $f$ modify the potential \eqref{eq:fullpotential} as follows:
\begin{equation}
    \label{eq:potential-corrections}
    V=V_0\left(1 -\cloop\, b \left[\frac{1}{\phi^{2/3}} + \mathfrak{a} + \mathfrak{b} ~\phi^{2/3} +\dots\right]\right)\,.
\end{equation}
Here we introduced constants $\mathfrak{a},~\mathfrak{b}$ which generically should be $\O(1)$. The factor $\frak{a}$ only influences the height of the inflationary plateau. This effect is negligible due to the smallness of the coefficient $(b\, \cloop)$ and we will disregard $\frak{a}$ in what follows. Depending on sign and value of $\frak{b}$ (as well as possibly of prefactors of higher terms in our expansion) we could find whole classes of models of slow-roll inflation. However, for reasons that will become clear shortly we will include only the first relevant correction $\sim \frak{b} \phi^{2/3}$ for the following analysis.

The slow-roll parameters for the potential \eqref{eq:potential-corrections} read:
\begin{align}
    & \epsilon  \simeq \frac{1}{2} (b\, \cloop)^2 \left[ \frac49 \frac{1}{\phi^{10/3}}- \frac89 \frac{\mathfrak{b}}{\phi^{6/3}} + \frac49 \frac{\frak{b}^2}{ \phi^{2/3}}\right], \\
    & \eta  \simeq - b\, \cloop \left[ \frac{10}{9}\frac{1}{\phi^{8/3}}- \frac 29 \frac{\frak{b}}{\phi^{4/9}} \right]. 
 \end{align}
We see that slow roll is possible due to the small prefactor $(b\,\cloop)$.
Using these results we find the number of efoldings:
\begin{equation}
    \label{eq:Ne-more-models-full}
    N_e = \int_{\phi_{\rm end}}^{\phi_*} \frac32\frac{\phi^{5/3}}{b \cloop \left[ 1 -  \mathfrak{b}\, \phi^{4/3}\right]}\, \dd \phi\,.
\end{equation}
Again, the exact number of $N_e$ is determined by the post-inflationary history and we treat $N_e$ as a constant such that we can read \eqref{eq:Ne-more-models-full} as a constraint which determines~$\phi_*$.
The integral \eqref{eq:Ne-more-models-full} is dominated by the largest $\phi$-values such that we may replace the lower integration limit by zero.
Assuming $\mathfrak{b}>0$, we see that 
compared to the situation with $\mathfrak{b}=0$ a smaller value of $\phi_*$ will be sufficient to give a predetermined value of $N_e$.
This is encouraging since it indicates that the presence of the $\mathfrak{b}$-correction actually leads to a more robust scenario for inflation and we are not required to go to very large values of $\phi_*$.

We now want to proceed with the following logic: First, we recall that our analysis without subleading corrections gave $\phi_*\simeq 0.2$. The smallness of this value relative to unity may be viewed as a result of the smallness of the power spectrum. We may think of $\phi_*$ as of the small parameter of our analysis. As we just argued, this value only becomes smaller for non-zero $\frak{b}$. Thus, we may treat $\frak{b}\phi^{4/3}$ in \eqref{eq:Ne-more-models-full} 
as a small correction and evaluate the integral perturbatively:
\begin{equation}
    \label{eq:Ne-more-models}
    N_e\simeq \frac{9}{16} \frac{\phi_*^{8/3}}{b\,\cloop} (1 + 2\frak{b} \phi_*^{4/3})\,.
\end{equation}
This confirms that the correction tends to make $\phi_*$ smaller for given $N_e$.
In addition, we want to analyse the effects induced by the normalization of scalar perturbations,
\begin{equation}
    \label{eq:COBE-more-models}
    \hat{A}_s = \frac{9V_0}{4 (b\, \cloop)^2} \phi_*^{10/3}\left(1+2 \frak{b}\phi^{4/3}\right),
\end{equation}
where we used the approximation $1-\cloop b (\phi_*^{-2/3}+ \frak{b} \phi_*^{2/3})\simeq 1$ and expanded to leading order in $\frak{b} \phi_*^{4/3}$.
Recall that the measured value for $\hat{A}_s$ is given by \eqref{eq:scalar-perturbations-potential}.

Analogously to Sec.~\ref{sc:para-control}, we now solve \eqref{eq:Ne-more-models} and \eqref{eq:COBE-more-models} for $\phi_*$ and $\V$ where we have to keep in mind that the volume $\vol$ is contained in $V_0$ and $b$, see \eqref{eq:V0} and \eqref{eq:fullpotential}.
We can use \eqref{eq:Ne-more-models} and solve for $\V$ in terms of $\phi_*$
\begin{equation}
    \label{eq:volume-sol-more-models}
    \V=\frac{A}{\phi^{8}_*} \left(1+2\frak{b} \phi^{4/3}_*\right)^{-3}\,.
\end{equation}
Using this relation for $\V$ in \eqref{eq:COBE-more-models} we obtain the following equation:
\begin{equation}
\label{eq:phi-sol-more-models}
    \left( B A^7 \right)^{2/99} = \phi_*^{4/3} \left(1+2\frak{b} \phi^{4/3}_*\right)^{48/99}\,.
\end{equation}
For $\mathfrak{b}=0$, we find that \eqref{eq:volume-sol-more-models} and \eqref{eq:phi-sol-more-models} are equivalent to \eqref{eq:volume-sol-simple} and \eqref{eq:phi-sol-simple} respectively.
To leading order in $\mathfrak{b}\phi_*^{4/3}$, \eqref{eq:phi-sol-more-models} is a quadratic equation for $\phi_*^{4/3}$ which is solved by
\begin{equation}
    \label{eq:sol-for-phi-more-models}
    \phi_*^{4/3} = \frac{33}{64 \frak{b}}(-1 \pm 1)\pm \left( B A^7 \right)^{2/99} \mp \frac{32}{33} \frak{b} \left( B A^7 \right)^{4/99} + \O\left(\frak{b}^2 \left( B A^7 \right)^{6/99}\right).
\end{equation}
The physical solution corresponds to the upper sign choice in \eqref{eq:sol-for-phi-more-models}:
\begin{equation}
    \label{eq:sol-for-phi-more-models-physical}
    \phi_* = \left( B A^7 \right)^{1/66}\left[ 1 - \frac{8}{11} \frak{b} \left( B A^7 \right)^{2/99}\right].
\end{equation}
We can now clearly see that the correction with $\frak{b}>0$ will lower $\phi_*$.
Furthermore, we obtain the solution for $\V$ by using \eqref{eq:sol-for-phi-more-models-physical} in \eqref{eq:volume-sol-more-models} and again expanding to leading order:
\begin{equation}
    \label{eq:sol-for-vol-more-models-final}
    \V = \left(\frac{A^5}{B^4}\right)^{\frac{1}{33}} \left( 1 - \frac29 \frak{b} \left(B A^7\right)^{2/99}\right).
\end{equation}
The results \eqref{eq:sol-for-phi-more-models-physical} and \eqref{eq:sol-for-vol-more-models-final} reproduce  \eqref{eq:solutions-NQ} by setting~\mbox{$\mathfrak{b}=0$}. We see that not only $\phi_*$ but also $\V$ are lowered for~$\mathfrak{b}>0$, though the effect on the volume is weaker.

At this stage, it becomes clear that a variety of different models can arise when the corrections in \eqref{eq:corrections-in-f} are taken into account. We also see that, if the sign of the prefactor $\frak{b}$ turns out right, the leading correction allows for more robust model thanks to the smaller value of $\phi_*$. It may be interesting to further investigate these scenarios, including higher terms from \eqref{eq:corrections-in-f}. However, a posteriori the lower value of $\phi_*$ can justify our neglect of such sub-leading corrections. Moreover, it also lends support to the simplest version of our analysis in Sec.~\ref{sc:simplest-setting}, where even the leading correction $\sim\frak{b}$ was disregarded.

\subsection{Original blow-up regime}
\label{sc:Original-blow-up-regime}
In this section we determine the critical value of $\cloop$ for which the original blow-up model~\cite{Conlon:2005jm} transitions to Loop Blow-up inflation. A parametric estimate demonstrating that loop corrections tend to destroy slow roll in blow-up inflation appears in [\citealp{Cicoli:2008gp}, \citealp{Baumann:2014nda}]. Yet, the critical value of the prefactor $c_{\rm loop}$ has not been derived. Note also that the effect of higher derivative $\alpha'^3$ effects has been studied in \cite{Cicoli:2023njy}. The result is that, for the expected value of their prefactor, they do not spoil the flatness of the inflationary plateau and can instead improve the agreement of the scalar spectral index with CMB data.

To determine a critical value for $\cloop$, imagine we can treat $\cloop$ as a free parameter which we set to zero initially.
In this setting, we implement all our phenomenological constraints on $N_e$, $n_s$ and the normalization of scalar perturbations, thereby fixing some of the parameters of the blow-up inflation model.
Now we increase $\cloop$, insisting that the model is not significantly affected.
In particular, we demand that the relative corrections $\delta\eta/\eta$ and $\delta\epsilon/\epsilon$ (and hence the correction to $N_e$) remain small.
This will determine a critical value for $\cloop$.
To obtain $\delta\epsilon/\epsilon$ and $\delta\eta/\eta$, we first rewrite the potential \eqref{eq:tauphi-pot} as:
\begin{equation}
V(\phi) = V_0 + V_\textrm{np}(\phi) + V_\textrm{loop}(\phi)\,,
\end{equation}
where we defined
\begin{gather}
\label{eq:Vbu-and-Vloop-def}
V_\textrm{np}(\phi)= -V_0 \B_\phi\frac{{\V}}{\beta} \, \tau_\phi(\phi) \, e^{-a_\phi\tau_\phi(\phi)} \,,\qquad
V_\textrm{loop}(\phi)= - V_0 \frac{c_\text{loop}}{\beta\sqrt{\tau_\phi(\phi)}}\,,
\end{gather}
with $V_\textrm{np}(\phi)$ corresponding to the non-perturbative potential which generates slow-roll inflation in \cite{Conlon:2005jm}.
Assuming, as explained, $V_\textrm{loop}'\ll V_\textrm{np}'$, we have:
\begin{eqnarray}
    \epsilon &\simeq& \frac12 \left(\frac{V_\textrm{np}'(\phi) + V_\textrm{loop}'(\phi)}{V_0}\right)^2= \frac12 \left(\frac{V_\textrm{np}'(\phi) }{V_0}\right)^2 \left(1 + 2\frac{V_\textrm{loop}'(\phi)}{V_\textrm{np}'(\phi)}+\dots \right), \\
    \eta &\simeq&  \frac{V_\textrm{np}''(\phi) }{V_0} \left(1 +\frac{V_\textrm{loop}''(\phi)}{V_\textrm{np}''(\phi) } \right).
\end{eqnarray}
Using \eqref{eq:Vbu-and-Vloop-def}, the relative corrections become
\begin{align}
    \label{eq:delta-epsilon}
    \frac{\delta\epsilon}{\epsilon} &\simeq 2\frac{V_\textrm{loop}'(\phi)}{V_\textrm{np}'(\phi)}= \frac{ \cloop \tau_\phi^{-3/2}}{\B_\phi \vol (a_\phi \tau_\phi - 1) e^{-a_\phi \tau_\phi}}\,,\\[5pt]
    \frac{\delta\eta}{\eta} &\simeq \frac{V_\textrm{loop}''(\phi)}{V_\textrm{np}''(\phi)}
    =\frac58 \frac{\cloop \tau_\phi^{-3/2}}{\B_\phi \vol ~a_\phi \tau_\phi ~e^{-a_\phi \tau_\phi}}\left[ a_\phi \tau_\phi-\frac94+(4a_\phi\tau_\phi)^{-1}\right]^{-1} \,.
\end{align}
The ratio of these corrections at horizon crossing, $\tau_\phi=\tau_{\phi_*}$, is:
\begin{equation}
    \frac{\delta \epsilon/\epsilon}{ \delta \eta/\eta} = \frac85 a_\phi \tau_{\phi_*} ~\frac{ a_\phi \tau_{\phi_*}-\frac94 + (4a_\phi \tau_{\phi_*})^{-1}}{a_\phi \tau_{\phi_*}-1} \gg 1\,,
\end{equation}
since inflation takes place in a regime where $a_\phi\tau_{\phi_*} > a_\phi\left< \tau_\phi \right>\simeq \ln \vol \gg 1$. Thus, the main correction we need to control is $\delta \epsilon/\epsilon$. Recall that, while the spectral index is dominated by $\eta$, the parameter $\epsilon$ is nevertheless essential since it governs the number of efoldings $N_e$. 

To estimate whether the correction is significant we need to determine $\tau_{\phi_*}$ and $\V$. In the original blow-up inflation model, the number of efoldings $N_e$ is given by (cf. (4.35) in~\cite{Cicoli:2016olq}):
\begin{equation}
    \label{eq:Ne}
    N_e = \frac{\kappa_e}{\vol^2}\,\frac{e^{a_\phi \tau_{\phi_*}}}{(a_\phi \tau_{\phi_*})^{3/2}}\,,\qquad \kappa_e\equiv \frac{3 \beta W_0 \lambda_\phi}{16 a_\phi^{3/2} A_\phi}\,,
\end{equation}
and the normalization of scalar perturbation reads (cf. (4.39) in \cite{Cicoli:2016olq}):
\begin{equation}
    \label{eq:norm}
 \hat{A}_s = \frac{\kappa_s}{\sqrt{a_\phi \tau_{\phi_*}}(a_\phi \tau_{\phi_*} -1)^2} \frac{e^{2a_\phi\tau_{\phi_*}}}{\vol^6}\,,\qquad \kappa_s\equiv \left( \frac{\gs e^{K_{\rm cs}}}{8\pi}\right) \frac{3 \lambda_\phi \beta^3 W_0^2}{64 a_\phi^{3/2}} \left(\frac{W_0}{ A_\phi} \right)^2\,.
\end{equation}
Recall that the measured value for $\hat{A}_s$ is given by \eqref{eq:scalar-perturbations-potential}. We can use \eqref{eq:Ne} and \eqref{eq:norm} to solve for $\tau_{\phi_*}$ and $\vol$.
In a first step we use \eqref{eq:Ne} to solve for $\vol$ in terms of $\tau_{\phi_*}$
\begin{equation}
    \label{eq:vol-exponential}
    \vol = \left( \frac{\kappa_e}{N_e} \frac{e^{a_\phi \tau_{\phi_*}}}{(a_\phi \tau_{\phi_*})^{3/2}}\right)^{1/2}\,,
\end{equation}
such that we can eliminate $\vol$ in \eqref{eq:norm}. Suitably rewriting \eqref{eq:norm} in the limit $a_\phi \tau_{\phi_*}\gg 1$ one finds
\begin{equation}
    \label{eq:exponentials}
    (a_\phi \tau_{\phi_*})^{-2} e^{a_\phi \tau_{\phi_*}}=\left( \frac{N_e}{\kappa_e} \right)^3 \frac{\kappa_s}{\hat{A}_s}~,
\end{equation}
which is approximately solved by
\begin{equation}
    \label{eq:approx-sol}
    a_\phi \tau_{\phi_*} \simeq \ln\left[\left( \frac{N_e}{\kappa_e} \right)^3 \frac{\kappa_s}{\hat{A}_s} \right]+ \O(1)\,.
\end{equation}
Then, using \eqref{eq:vol-exponential} and \eqref{eq:approx-sol} in \eqref{eq:delta-epsilon} and reinstating the factors in $\kappa_e$ and $\kappa_s$ gives:
\begin{equation}
\label{eq:explicit-delta-epsilon}
    \frac{\delta \epsilon}{\epsilon} \simeq \cloop   \sqrt{\frac{8}{27\pi}\frac{e^{K_{\rm cs} }\gs}{\hat{A}_s \beta}}
    \,\frac{N_e^2 W_0 a_\phi^{5/4}}{\lambda_\phi^{3/2}} \left( \ln\left[ \frac{8 a_\phi^3 A_\phi e^{K_{\rm cs}} \gs N_e^3 W_0}{9\pi \hat{A}_s \lambda_\phi^2}\right]\right)^{-3/4}\,.
\end{equation}
We can evaluate \eqref{eq:explicit-delta-epsilon} for the typical parameter choice (\ref{Parameters}) together with $a_\phi=2\pi$, $A_\phi=1$ and $N_e=52$ which yields
\begin{equation}
    \delta \epsilon/\epsilon \simeq 2.4 \times 10^6\,\cloop\,.
\end{equation}
Thus, in order to avoid considerable deviations from the predictions for the main cosmological observables, we have to demand:
\begin{equation}
\label{eq:upper bound on cloop}
\delta \epsilon/\epsilon\ll 1\qquad\Leftrightarrow\qquad    \cloop \ll 0.4 \times10^{-6}\,.
\end{equation}
This result has been checked via a detailed numerical analysis by computing the values of $\phi_*$ and $\V$ yielding $N_e \simeq 52$ (in agreement with the values of $N_e$ that we will obtain in Sec. \ref{sc:pheno}) while ensuring the correct normalization of scalar perturbations based on~\eqref{eq:norm}. We then turn on loop corrections and increase the value of $\cloop$ until their contribution spoils the predictions of the original model. Generically, the coefficient $\cloop$ is expected to be small due to a suppression by factors of $2\pi$. A careful analysis in \cite{Gao:2022uop} estimates $\cloop\sim(2\pi)^{-4}\sim10^{-4}$. Indeed, this value could suffice to neglect loop corrections if $\gs$ and $W_0$ are tuned appropriately small in \eqref{eq:explicit-delta-epsilon}. However, the smallness of $g_s$ is limited  by the fact the volume is exponentially large in $1/g_s$. Moreover, tuning $W_0$ to a small value goes together with making the volume small. This is, in turn, highly problematic because of warping corrections, as discussed in detail in \cite{Junghans:2022exo, Gao:2022fdi} and also in Sec.~\ref{sc:warping-corr}. Thus, we conclude that loop corrections tend in general to spoil the original model of blow-up inflation based purely on non-perturbative effects.

\section{Phenomenological analysis}
\label{sc:pheno}

In this section we want to derive precise predictions for the cosmological parameters of our model. To accomplish this, we have to study the post-inflationary evolution. Following the analysis in \cite{Cicoli:2023njy}, we will obtain the number of efoldings of inflation $N_e$, from which all the other parameters follow.

\subsection{Moduli decay rates and dark radiation}

Reheating is the process through which the inflationary energy is transferred into the Standard Model (SM). To understand this process, it is therefore crucial to first identify the location of the SM in the extra-dimensions, and subsequently to compute all the moduli couplings and decay rates to the SM as well as hidden degrees of freedom.

\subsubsection*{Standard Model realization}

As pointed out in \cite{Blumenhagen:2007sm}, it is very hard to stabilize the SM cycle via non-perturbative effects since chiral intersections between instantons and SM matter fields tend to give a vanishing prefactor of the non-perturbative contributions to $W$. Since we require non-perturbative effects to generate a minimum at the end of inflation, we will not realize the SM on D7-branes wrapped around $\tau_\phi$.\footnote{Even a loop stabilization of $\tau_\phi$ would not work since non-zero gauge fluxes needed for chiral matter would generate a $\tau_\phi$-dependent Fayet-Iliopoulos term that would spoil the flatness of the inflationary plateau.} Instead, we will include an additional blow-up cycle, denoted as $\tau_\SM$. If $\tau_\SM$ is in the geometric regime, the SM can live on D7-branes wrapped around this divisor. If, by contrast, $\tau_\SM$ collapses to zero size, the SM is realized on fractional D3-branes. The total internal volume therefore takes the form:
\begin{equation}
\label{eq:volume with tau_SM}
    \V = \tau_b^{3/2}- \lambda_s \tau_s^{3/2}- \lambda_\phi \tau_\phi^{3/2}- \lambda_\SM \tau_\SM^{3/2}-\lambda_{\rm int}\left(\tau_{\rm int}-\lambda \tau_\SM\right)^{3/2}\,,
\end{equation}
where, on top of $\tau_\SM$, we included another divisor $\tau_{\rm int}$ which intersects with $\tau_\SM$. Let us describe the stabilization of $\tau_\SM$ following the discussion in \cite{Cicoli:2012sz}. Non-zero gauge fluxes on the D7-stack wrapped around $\tau_\SM$ generate chiral matter and a moduli-dependent Fayet-Iliopoulos term. For zero VEVs of the charged matter fields, D-term stabilization yields a vanishing Fayet-Iliopoulos term which, for an appropriate choice of gauge fluxes, corresponds to \cite{Cicoli:2012sz}:
\begin{equation}
\tau_\SM = \left(\frac{\lambda_{\rm int}\lambda}{\lambda_\SM}\right)^2 \left(\tau_{\rm int}-\lambda \tau_\SM\right).
\label{Dstab}
\end{equation}
For geometries without an intersecting divisor, i.e. with $\lambda_{\rm int}=0$, (\ref{Dstab}) leads to $\tau_\SM\to 0$, forcing this 4-cycle to shrink down to zero size. In this case the SM would live on D3-branes at a CY singularity. When instead $\lambda_{\rm int}\neq 0$, (\ref{Dstab}) leaves a flat direction without forcing the collapse of any divisor. This direction can be parameterized by $\tau_\SM$ and can be fixed by string loops. Inspired by the explicit CY construction of \cite{Cicoli:2011qg}, the relevant loop potential might take the form: 
\begin{equation}
V(\tau_\SM) = \left(\frac{d_{\rm loop}}{\sqrt{\tau_\SM}}-\frac{g_{\rm loop}}{\sqrt{\tau_\SM}-\sqrt{\tau_s}}\right)\frac{W_0^2}{\vol^3}\,.
\end{equation}
It is easy to see that this potential admits a minimum at:
\begin{equation}
\tau_s = \left(1+\sqrt{\frac{g_{\rm loop}}{d_{\rm loop}}}\right)^2 \tau_\SM \sim \tau_\SM\,,
\end{equation}
showing that loops can fix the SM modulus $\tau_\SM$ in terms of $\tau_s$ which, in turn, is stabilized by non-perturbative effects. This allows to reproduce the 
correct SM gauge coupling $g_\SM^{-2}\simeq \tau_\SM \sim \tau_s \sim \mathcal{O}(10)$ for SM fields living on a D7-stack wrapped around $\tau_\SM$. 

\subsubsection*{Moduli decay rates}

The post-inflationary evolution is determined by the moduli decay rates. The only relevant moduli are the inflaton and the volume mode since $\tau_s$ and $\tau_\SM$ never come to dominate the energy density. The masses of the canonically normalized inflaton $\phi$ and volume $\chi$ read:
\begin{equation}
\label{eq:moduli masses}
 m_\phi  \simeq \frac{W_0 \ln \V}{\V}\,M_p\qquad\text{and}\qquad m_\chi  \simeq \frac{W_0}{\V^{3/2} \sqrt{\ln \V}}\, M_p\,, 
\end{equation}
where we introduced the reduced Planck mass $M_p\simeq 2.4\times 10^{18}$ GeV. Let us list the main decay rates for each modulus separately.
\begin{itemize}
\item \textbf{Volume $\chi$:} One of the leading decay channels of the volume mode is into its corresponding closed string axions $a_b$ with decay width \cite{Cicoli:2012aq,Higaki:2012ar}:
\begin{equation}
\Gamma_{\chi \to a_b a_b} = \frac{1}{48\pi}\frac{m_\chi^3}{M_p^2} \simeq \left(\frac{W_0^3}{48\pi \left(\ln\vol\right)^{3/2}}\right) \frac{M_p}{\vol^{9/2}}\,.
\label{GammaVolaxions}
\end{equation}
Moreover, logarithmic loop corrections to the Higgs mass induce a coupling between $\chi$ and SM Higgs scalars $h$ with decay rate \cite{Cicoli:2022fzy}: 
\begin{equation}
\label{eq:enhanced decay rate V to hh}
    \Gamma_{\chi \to hh} = \frac{\tilde c_{\rm loop}^2}{32\pi} \left(\frac{m_0}{m_\chi}\right)^4\frac{m_\chi^3}{M_p^2}\,,
\end{equation}
where $\tilde c_{\rm loop} \simeq 1/(16 \pi^2) \sim \O(10^{-2})$ is a 1-loop factor and $m_0$ the soft SUSY breaking scalar mass. These two parameters determine the ratio between (\ref{GammaVolaxions}) and (\ref{eq:enhanced decay rate V to hh}):
\begin{equation}
\frac{\Gamma_{\chi \to hh}}{\Gamma_{\chi \to a_b a_b}} \simeq \tilde c_{\rm loop}^2 \left(\frac{m_0}{m_\chi}\right)^4\,.
\end{equation}
When the SM lives on D7-branes, $m_0\simeq m_{3/2}\simeq W_0 M_p/\vol \gg m_\chi$ \cite{Conlon:2006wz}, while when the SM is sequestered on D3-branes, $m_0\lesssim m_\chi$ \cite{Blumenhagen:2009gk, Aparicio:2014wxa}, implying for $\vol \sim 10^4$: 
\begin{eqnarray}
\frac{\Gamma_{\chi \to hh}}{\Gamma_{\chi \to a_b a_b}} &\simeq&  \left( \tilde c_{\rm loop} \vol\right)^2 \gg 1 \qquad\text{for SM on D7} 
\label{RatioD7} \\
\frac{\Gamma_{\chi \to hh}}{\Gamma_{\chi \to a_b a_b}} &\lesssim&  \tilde c_{\rm loop} \ll 1 \qquad\quad\,\,\,\,\,\text{for SM on D3}
\end{eqnarray}
Hence, when the SM is on D7-branes, (\ref{eq:enhanced decay rate V to hh}) dominates over (\ref{GammaVolaxions}). Plugging \eqref{eq:moduli masses} into~\eqref{eq:enhanced decay rate V to hh} we get an expression of the decay rate of the volume mode in terms of $\V$ only:
\begin{equation}
\label{eq:decay rate of V}
    \Gamma_{\chi\to hh} \simeq \left(\frac{\tilde c_{\rm loop}^2\, W_0^3 \,\sqrt{\ln\vol}}{32\pi}\right)\frac{M_p}{\vol^{5/2}}\,.
\end{equation}
On the other hand, when the SM in on D3-branes, the decay rate (\ref{eq:enhanced decay rate V to hh}) can be safely ignored. In this case, the main volume decay channel into SM degrees of freedom is induced by a Giudice-Masiero interaction in the K\"ahler potential between the volume mode and Higgs bosons $H_u$ and $H_d$ with coefficient $Z$. The corresponding decay rate is given by \cite{Cicoli:2012aq}:
\begin{equation}
\Gamma_{\chi \to H_u H_d} = \frac{Z^2}{24\pi}\frac{m_\chi^3}{M_p^2} \simeq \left(\frac{Z^2\,W_0^3}{24\pi \left(\ln\vol\right)^{3/2}}\right) \frac{M_p}{\vol^{9/2}}\,.
\label{GMDecay}
\end{equation}

\item \textbf{Inflaton $\phi$:} When the inflaton 4-cycle is wrapped by a hidden D7-stack, the main decay rate of $\phi$ is into light hidden sector gauge bosons $\gamma_h$ and looks like \cite{Conlon:2007gk}:
\begin{equation}
\Gamma_{\phi\to\gamma_h\gamma_h} \simeq \frac{\vol}{64\pi}\frac{m_\phi^3}{M_p^2} \simeq 
\left(\frac{\left(W_0 \,\ln \V\right)^3}{64\pi}\right)\frac{M_p}{\vol^2}\,.
\label{eq:Gamma phi to 2gamma_h}
\end{equation}
On the other hand, the situation when the inflaton divisor is not wrapped by any D7-brane has been studied in \cite{Cicoli:2022fzy}. The main two-body inflaton decay channels are into volume moduli $\chi$ and volume axions $a_b$ and scale as:
\begin{equation}
\Gamma_{\phi\to \chi\chi}\simeq  \Gamma_{\phi\to a_b a_b}
\simeq \frac{\left(\ln\vol\right)^{3/2}}{64\pi\,\vol}\frac{m_\phi^3}{M_p^2}\simeq \left(\frac{W_0^3 \left(\ln\vol\right)^{9/2}}{64\pi}\right)\frac{M_p}{\vol^4}\,.
\label{InflatonDecays}
\end{equation}
After being produced from the inflaton decay, the volume moduli behave as\linebreak explained above, depending on the realization of the SM. When the SM is on D7-branes, $\chi$ decays almost instantaneously into SM Higgses $h$ since from (\ref{eq:decay rate of V})\linebreak $\Gamma_{\chi\to hh}\sim \tilde c_{\rm loop}^2 M_p / \vol^{5/2} \gg \Gamma_{\phi \to \chi\chi}$ for $\vol \gg 1$. When instead the SM is on D3-branes, comparing~(\ref{GMDecay}) with (\ref{InflatonDecays}) for large $\vol$, we realize that $\Gamma_{\chi\to H_u H_d}\ll \Gamma_{\phi\to \chi\chi}$, implying that $\chi$ decays after $\phi$ diluting all the inflaton decay products. 

Moreover, when the SM lives on D7-branes, there are additional inflaton decay channels which scale as (\ref{InflatonDecays}). These are $\phi$-decays into pairs of SM gauge bosons $\gamma$, SM axions $a_\SM$ (which play the role of QCD axions) and SM moduli $\tau_\SM$. Given that $\Gamma_{\tau_\SM\to \gamma\gamma}\simeq \Gamma_{\tau_\SM\to a_\SM a_\SM}\sim M_p/\vol^2$ is much larger than (\ref{InflatonDecays}) for $\vol \gg 1$, after being produced from the inflaton decay, $\tau_\SM$ decays almost instantaneously in $\gamma\gamma$ and $a_\SM a_\SM$ with \cite{Cicoli:2022fzy}:
\begin{equation}
\frac{\Gamma_{\tau_\SM \to \gamma\gamma}}{\Gamma_{\tau_\SM\to a_\SM a_\SM}} = 8 \,N_g \geq 96 \gg 1\,,
\end{equation}
where we have considered a number of gauge bosons $N_g \geq 12$ which is at least as big as in the SM. 
\end{itemize}

\subsubsection*{Dark radiation}

As we have just seen, the decays of $\phi$ and $\chi$, besides producing SM particles, yield very light axions, like $a_b$ and $a_\SM$, which are relativistic and can contribute to extra dark radiation~\cite{Cicoli:2012aq,Higaki:2012ar,Hebecker:2014gka,Angus:2014bia,Allahverdi:2014ppa,Cicoli:2015bpq,Cicoli:2022fzy,Cicoli:2022uqa, Allahverdi:2013noa,Cicoli:2018cgu}. This is parameterized by $\Delta N_{\rm eff}$, the effective number of additional neutrino-like species with respect to the SM case. It may happen that the axions produced in the decay of the heaviest modulus do not contribute to dark radiation because they are diluted by the decay of the lightest modulus. This situation arises if the latter comes to dominate the energy density before decaying. In this case, the axionic contribution to $\Delta N_{\rm eff}$ is determined by the decay of the lightest modulus which we denote as $\sigma$. Writing the decay of $\sigma$ into SM particles as $\Gamma_{\sigma\to\SM}$, and as $\Gamma_{\sigma \to {\rm hid}}$ the $\sigma$-decay into hidden degrees of freedom which we assume to be just closed string axions, the axionic contribution to extra dark radiation can be computed as \cite{Cicoli:2012aq, Higaki:2012ar}:
\begin{equation}
\Delta N_{\rm eff} = \frac{43}{7}\frac{\Gamma_{\sigma \to {\rm hid}}}{\Gamma_{\sigma \to \SM}}\left(\frac{10.75}{g_*(T_{\rm rh})}\right)^{1/3}\,,    
\label{DNeff}
\end{equation}
where $g_*(T_{\rm rh})$ is the number of relativistic degrees of freedom at the reheating temperature~$T_{\rm rh}$. This prediction has to be confronted with constraints from CMB observations which set a tight upper bound on $\Delta N_{\rm eff}$ (depending on the specific dataset used) \cite{Planck:2018vyg}:
\begin{equation}
\label{eq:Planck constraint on Delta Neff}
    \Delta N_{\rm eff} \lesssim 0.2-0.5\quad\text{at}\,\,95\%\,\,\text{CL}\,.
\end{equation}
Let us now use the moduli decay rates computed above to evaluate $\Delta N_{\rm eff}$ for the following different scenarios:
\begin{itemize}
\item[\it I)] \textit{SM on D7s and inflaton wrapped by D7s:} In this case $\sigma\equiv \chi$, and so $\Gamma_{\sigma \to {\rm hid}}$ is given by (\ref{GammaVolaxions}), while $\Gamma_{\sigma \to \SM}$ is given by (\ref{eq:enhanced decay rate V to hh}). Plugging (\ref{RatioD7}) into (\ref{DNeff}), it turns out that $\Delta N_{\rm eff}\simeq 0$ since the volume mode decays predominantly into SM Higgses.

\item[\it II)] \textit{SM on D7s and inflaton not wrapped by any D7:} In this case $\sigma\equiv \phi$, $\Gamma_{\sigma \to \SM} = \Gamma_{\phi \to \chi\chi \to hhhh} + \Gamma_{\phi \to \gamma\gamma} + \Gamma_{\phi \to \tau_\SM\tau_\SM \to \gamma\gamma\gamma\gamma}$, while $\Gamma_{\sigma\to{\rm hid}} = \Gamma_{\phi\to a_b a_b} + \Gamma_{\phi \to a_\SM a_\SM} + \Gamma_{\phi \to \tau_\SM\tau_\SM \to a_\SM a_\SM a_\SM a_\SM}$. This case has been analyzed in detail in \cite{Cicoli:2022fzy} which found $\Delta N_{\rm eff}\simeq 0.14$ setting $N_g=12$ and $g_*(T_{\rm rh})=106.75$. 

\item[\it III)] \textit{SM on D3s:} When the SM is on D3s, the last modulus to decay is always $\chi$ regardless of the fact that the inflaton is wrapped or not by a D7-stack. Hence $\sigma\equiv \chi$ and $\Gamma_{\sigma \to {\rm hid}}$ is given by (\ref{GammaVolaxions}), but now $\Gamma_{\sigma \to \SM}$ is given by (\ref{GMDecay}). Plugging these results into (\ref{DNeff}), we find $\Delta N_{\rm eff} \simeq 1.43/Z^2$ for $g_*(T_{\rm rh})=106.75$ since we shall see that $T_{\rm rh}$ is well above the EW scale where all SM degrees of freedom are relativistic. Imposing $\Delta N_{\rm eff}\lesssim 0.5$ requires $Z\gtrsim 1.7$.
\end{itemize}

\subsection{Post-inflationary dynamics}
\label{sc:post-inflationary-dynamics}

We will analyze three different scenarios of post-inflationary evolution. In \textit{scenario I} the SM lives on D7-branes and the inflaton 4-cycle is wrapped by a stack of hidden-sector D7-branes, as considered in \cite{Cicoli:2010ha,Cicoli:2010yj,Allahverdi:2020uax}. On the other hand, \textit{scenario II} corresponds to the case envisaged in \cite{Cicoli:2022fzy} where the SM is realized again via D7-branes but the inflaton 4-cycle is not wrapped by any D7-branes
. Finally, in \textit{scenario III} we will analyze the case where the SM is on D3-branes at singularities, regardless of the presence of a D7-stack on the inflaton 4-cycle (for reheating from moduli decay in D3-models see \cite{Cicoli:2012aq, Higaki:2012ar, Allahverdi:2013noa, Cicoli:2022uqa, Cicoli:2015bpq}). 

\subsubsection*{Scenario I: Reheating from volume mode decay}

As already explained, in the case when the inflaton is wrapped by a hidden D7-stack and the SM is on D7-branes, the inflaton decays promptly into hidden sector degrees of freedom via (\ref{eq:Gamma phi to 2gamma_h}).\footnote{Here we are assuming that the hidden sector D7-stack on $\tau_\phi$ is not a pure SYM theory that develops a mass gap above $m_\phi$, as considered in \cite{Cicoli:2010ha}. In fact, in this case the decay of $\phi$ into hidden sector gauge bosons would be kinematically forbidden and the post-inflationary evolution would be the same as in the situation where $\tau_\phi$ is not wrapped by any D7.} These hidden particles are then diluted by the subsequent decay of the volume mode that leads to the final reheating via (\ref{eq:enhanced decay rate V to hh}). 

Let us now analyze in detail the post-inflationary evolution of this scenario. At the end of inflation, both the inflaton and the volume modulus are displaced from their post-inflationary minima and start oscillating. These oscillations redshift as \cor{matter}, and the energy density stored in the inflaton is larger than that in the volume mode. In fact, at the end of inflation, at time $t_{\rm end}$,  the energy density of the inflaton can be approximated as the inflation scale:
\begin{equation}
\label{eq:rho_phi(t_E)}
\rho_{\phi}(t_{\rm end}) \simeq 3\,H_{\rm inf}^2 M_p^2\simeq \frac{\beta W_0^2}{\V^3}\,M_p^4\,.
\end{equation}
On the other hand, the energy density stored in the volume modulus is:
\begin{equation}
\label{eq:rho_V(t_E)}
\rho_\chi(t_{\rm end}) \simeq m_\chi^2 \chi_0^2 \simeq \frac{W_0^2 \,Y^2}{\V^3 \ln \V}\,M_p^4\,,
\end{equation}
where $Y$ is the displacement in Planck units of the minimum of $\chi$ during inflation, i.e. $\chi_0= Y\,M_p$, which can be written as \cite{Cicoli:2016olq}:
\begin{equation}
    Y \simeq \sqrt{\frac{2}{3}} R \, (\ln \V)^{3/2}\,,
\end{equation}
with:
\begin{equation}
    R \equiv \frac{\lambda_\phi a_\phi^{-3/2}}{\lambda_s a_s^{-3/2}+ \lambda_\phi a_\phi^{-3/2}}\ll 1\,.
\end{equation}
In \cite{Cicoli:2016olq}, using $R \sim 0.1-0.01$, $Y$ was estimated to be of order $Y \sim 0.1$. Therefore, the ratio of the energy densities at the end of inflation is:
\begin{equation}
\label{eq:theta parameter}
    \theta\equiv \frac{\rho_\chi(t_{\rm end})}{\rho_{\phi}(t_{\rm end})} \simeq \frac{Y^2}{\beta \ln \V} \ll 1\,.
\end{equation}
Immediately after inflation, an era of matter domination starts, driven by the coherent oscillations of the inflaton field. This lasts until the inflaton decays at the time $t_{{\rm dec},\phi}$. We will indicate the number of efoldings of inflaton domination as $N_\phi$ which is given by:
\begin{equation} 
\label{eq:N_phi SI}
    N_\phi = \ln \left(\frac{a(t_{{\rm dec},\phi})}{a(t_{\rm end})}\right) = \frac{1}{3}\ln \left(\frac{\rho_\phi(t_{\rm end})}{\rho_\phi(t_{{\rm dec},\phi})}\right)\simeq \frac{2}{3} \ln \left(\frac{H_{\rm inf}}{\Gamma_{\phi\to\gamma_h\gamma_h}}\right) \simeq \frac{2}{3} \ln \left(\frac{64 \pi }{W_0^2(\ln \V)^3}\sqrt{\frac{\beta\vol}{3}}\right).
\end{equation}
where in the last equality we have substituted \eqref{eq:Gamma phi to 2gamma_h} and \eqref{eq:rho_phi(t_E)}. At this point, the radiation produced by the inflaton decay comes to dominate the energy density. Since the volume oscillations continue to redshift as matter, there will be a time of volume-radiation equality denoted as $t_{\rm eq}$. The energy density and the Hubble scale at volume-radiation equality can be obtained by imposing:
\begin{equation}
\label{eq:matter-radiation equality condition}
    \rho_\chi(t_{\rm eq}) = \rho_{\rm rad}(t_{\rm eq})\,,
\end{equation}
which yields:
\begin{equation}
\label{eq:matter-radiation equality}
    \rho_\chi (t_{{\rm dec},\phi}) \left(\frac{a(t_{{\rm dec},\phi})}{a(t_{\rm eq})}\right)^3 = \rho_{\rm rad} (t_{{\rm dec},\phi}) \left(\frac{a(t_{{\rm dec},\phi})}{a(t_{\rm eq})}\right)^4\,.
\end{equation}
Until the inflaton decays at $t_{{\rm dec},\phi}$, the ratio of the energy densities of the volume modulus and the inflaton is constant and equal to \eqref{eq:theta parameter}, since both of them redshift as matter. Later on, when the inflaton decays at $t_{{\rm dec},\phi}$, it suddenly transfers its energy into radiation so that we have $\rho_{\rm rad}(t_{{\rm dec},\phi})= \rho_\phi(t_{{\rm dec},\phi})$. Using this relation, we find $a(t_{{\rm dec},\phi})/a(t_{\rm eq}) = \theta$. Hence, $\rho_{\rm rad}(t_{\rm eq}) \simeq \rho_{\rm rad}(t_{{\rm dec},\phi})\, \theta^4$ and:
\begin{equation}
\label{eq:H(t_eq)}
    H(t_{\rm eq}) \simeq H(t_{{\rm dec},\phi})\, \theta^2 \simeq \Gamma_{\phi\to\gamma_h\gamma_h}\, \theta^2\,.
\end{equation}
Therefore, inserting \eqref{eq:Gamma phi to 2gamma_h} in \eqref{eq:H(t_eq)}, we obtain:
\begin{equation}
    H(t_{\rm eq}) \simeq \left(\frac{\left(W_0\,\ln \vol\right)^3\theta^2}{64 \pi }\right) \frac{M_p}{\vol^2}.
\label{Heq}
\end{equation}
Starting at $t_{\rm eq}$, $\rho_\chi$ becomes dominant and a second era of matter domination starts. Again, this epoch lasts until the decay of the volume modulus at the time $t_{{\rm dec},\chi}$. We can determine the number of efoldings $N_\chi$ of volume domination as:
\begin{equation}
\label{eq:N_V in SI}
    N_\chi \simeq \frac{2}{3} \ln \left(\frac{H(t_{\rm eq})}{\Gamma_{\chi\to hh}}\right) \simeq \frac{2}{3} \ln \left(\frac{Y^4\, \sqrt{\vol\ln\vol}}{2\left(\beta\, \tilde c_{\rm loop}\right)^2}\right),
\end{equation}
where we have used \eqref{eq:decay rate of V}, \eqref{eq:theta parameter} and (\ref{Heq}). The volume mode decays at $t_{{\rm dec},\chi}$, giving rise to a second epoch of radiation domination with reheating temperature $T_{\rm rh}$ that can be estimated as:
\begin{equation}
\label{eq:equation for reheating temperature}
T_{\rm rh} = \left(\frac{90}{\pi^2 g_*(T_{\rm rh})}\right)^{1/4} \sqrt{M_p\, \Gamma_{\chi\to hh}}\,.
\end{equation}

\subsubsection*{Scenario II: Reheating from inflaton decay}

As already mentioned, in the case where the inflaton divisor is not wrapped by any D7-brane and the SM lives on a D7-stack, the final reheating is driven by the decay of the inflaton. In fact, in this case $\phi$ decays mainly into SM gauge bosons with decay width \cite{Cicoli:2022fzy}:
\begin{equation}
\Gamma_{\phi\to\gamma\gamma}\simeq 8\, N_g\,\Gamma_{\phi \to \chi\chi}\,,
\label{phitogammagamma}
\end{equation}
where $\Gamma_{\phi\to \chi\chi}$ is given by (\ref{InflatonDecays}). On the other hand, $\chi$ decays into SM Higgses with decay rate $\Gamma_{\chi\to hh}$ given by (\ref{eq:decay rate of V}). Hence the ratio between the two widths scales as:
\begin{equation}
\label{eq:ratio of the decay rates SII}
    \frac{\Gamma_{\phi\to \gamma\gamma}}{\Gamma_{\chi\to hh}} \simeq \frac{4\,N_g\, (\ln \V)^4}{\tilde c_{\rm loop}^2 \vol^{3/2}}\simeq 10^3\,,
\end{equation}
for $\vol\simeq 10^4$, $\tilde c_{\rm loop} \simeq 1/(16\pi^2)$ and $N_g = 12$. Thus, the volume would decay after the inflaton but, as we shall show below, when $\chi$ is not dominating the energy density. Therefore this scenario does not feature any epoch of volume domination. Let us see this more in detail.  

The number of efoldings of inflaton domination is:
\begin{equation}
\label{eq:N_phi in SII}
N_\phi \simeq \frac{2}{3} \ln \left(\frac{H_{\rm inf}}{\Gamma_{\phi\to\gamma\gamma}}\right) \simeq  \frac{2}{3} \ln \left(\sqrt{\frac{\beta}{3}}\frac{8\pi\,\vol^{5/2}}{N_g W_0^2 \left(\ln \vol\right)^{9/2}}\right).
\end{equation}
Similarly, the relation between $H(t_{\rm eq})$ and $H(t_{{\rm dec},\phi})$ is the same as \eqref{eq:H(t_eq)} but now with $H(t_{{\rm dec},\phi})\simeq \Gamma_{\phi\to\gamma\gamma}$. Thus, $H(t_{\rm eq})$ becomes:
\begin{equation}
\label{eq:H(t_eq) SII}
    H(t_{\rm eq}) \simeq H(t_{{\rm dec},\phi})\,\theta^2 \simeq \Gamma_{\phi\to\gamma\gamma}\,\theta^2\,.
\end{equation}
Now, computing the ratio between \eqref{eq:H(t_eq) SII} and $H(t_{{\rm dec},\chi})\simeq \Gamma_{\chi\to hh}$, and using (\ref{eq:theta parameter}) \linebreak and~(\ref{eq:ratio of the decay rates SII}), we find:
\begin{equation}
\label{eq:ratio of H(t_eq) and Gamma_V SII}
    \frac{H(t_{\rm eq})}{H(t_{{\rm dec},\chi})} \simeq \frac{\Gamma_{\phi\to \gamma\gamma}}{\Gamma_{\chi\to hh}}\,\theta^2\simeq 10^3\,\frac{Y^4}{\left(\beta \ln \vol\right)^2}\simeq 10^{-4}\,,
\end{equation}
for $\vol\simeq 10^4$, $Y\simeq 0.1$ and $\beta\simeq 2$. This result implies that the volume mode decays well before reaching volume-radiation equality. Hence, the number of efoldings of volume domination is exactly zero:
\begin{equation}
\label{eq:N_V=0 in SII}
    N_\chi = 0\,.
\end{equation}
Since the volume mode decays when it is a subdominant fraction of the energy density, reheating is driven by the inflaton decay. The corresponding reheating temperature turns out to be:
\begin{equation}
T_{\rm rh} = \left(\frac{90}{\pi^2 g_*(T_{\rm rh})}\right)^{1/4} \sqrt{M_p\, \Gamma_{\phi\to \gamma\gamma}}\,.
\end{equation}

\subsubsection*{Scenario III: Standard Model on D3-branes}

When the SM is realized on D3-branes at the CY singularity at $\tau_\SM \to 0$, we have seen that the final reheating is due to the decay of the volume mode. Let us now analyze in detail the post-inflationary evolution of this scenario depending on the presence or absence of a D7-stack wrapped on $\tau_\phi$.
\begin{itemize}
\item[\it III\,a)] \textit{Inflaton wrapped by D7s:} In this case the inflaton behaves as in scenario I, and so after the end of inflation decays very quickly into hidden sector gauge bosons via \eqref{eq:Gamma phi to 2gamma_h}. The number of efoldings of inflaton domination is therefore still given by (\ref{eq:N_phi SI}). The Hubble scale at volume-radiation equality is also unchanged and it is (\ref{Heq}). The modulus $\chi$ has instead a different behavior with respect to scenario I since it decays now later via (\ref{GMDecay}). Thus, the number of efoldings of volume domination becomes:
\begin{equation}
N_\chi \simeq \frac{2}{3} \ln \left(\frac{H(t_{\rm eq})}{\Gamma_{\chi\to H_u H_d}}\right) \simeq \frac{2}{3} \ln \left(\frac{3\,Y^4\, \left(\vol\ln\vol\right)^{5/2}}{8\left( \beta\, Z\right)^2}\right).
\end{equation}
The final reheating temperature associated to the decay of the volume mode looks like:
\begin{equation}
T_{\rm rh} = \left(\frac{90}{\pi^2 g_*(T_{\rm rh})}\right)^{1/4} \sqrt{M_p\, \Gamma_{\chi\to H_u H_d}}\,.
\label{TrhVol}
\end{equation}

\item[\it III\,b)] \textit{Inflaton not wrapped by any D7:} This time the dominant inflaton decay channels are into a pair of $\chi$ moduli and into a pair of $a_b$ axions with decay rate given by \eqref{InflatonDecays}, while the relevant volume decay rate is again \eqref{GMDecay}. Since $\Gamma_{\phi\to a_b a_b}/\Gamma_{\chi \to H_u H_d} \simeq \left(3/(8Z^2)\right) (\ln \V)^6 \sqrt{\V} \simeq 10^7$ for $Z\simeq 2$ and $\vol\simeq 10^4$, the inflaton decays before the volume. The number of efoldings of inflaton domination is:
\begin{equation}
N_\phi \simeq \frac{2}{3} \ln \left(\frac{H_{\rm inf}}{\Gamma_{\phi \to a_b a_b}}\right) \simeq \frac{2}{3} \ln \left(\sqrt{\frac{\beta}{3}}\frac{64 \pi \,\vol^{5/2}}{W_0^2 (\ln \V)^{9/2}}\right)\,.
\end{equation}
The inflaton decay products are relativistic and redshift as radiation even if they do not reach thermal equilibrium due to their feeble gravitational couplings. Their energy density becomes comparable to the one of the non-relativistic $\chi$ particles produced from the oscillations of the volume mode at:
\begin{equation}
H(t_{\rm eq}) \simeq H(t_{{\rm dec},\phi})\,\theta^2 \simeq \Gamma_{\phi\to a_b a_b}\,\theta^2\,.
\label{eq:H(t_eq)New}
\end{equation}
For our choice of parameters, this is also the energy scale when the $\chi$ particles produced from the inflaton decay become non-relativistic, at time $t_{\rm nr}$, since (denoting the momentum of $\chi$ particles as $p_\chi$):
\begin{equation}
p_\chi(t_{\rm nr}) =\frac{m_\phi}{2} \left(\frac{a(t_{{\rm dec},\phi})}{a(t_{\rm nr})}\right)\simeq m_\chi\quad\Rightarrow\quad \frac{a(t_{\rm eq})}{a(t_{\rm nr})}\simeq \frac{2}{\theta}\frac{m_\chi}{m_\phi}\simeq 
\frac{2 \beta}{Y^2}\frac{1}{\sqrt{\mathcal{V}\ln\mathcal{V}}}\simeq 1\,,
\end{equation}
for $\mathcal{V}\simeq 10^4$, $Y\simeq 0.1$ and $\beta\simeq 2$. This ensures that the $\chi$ particles produced from the inflaton decay indeed redshift as radiation until $t_{\rm eq}$.
Hence, the number of efoldings of the volume dominated era can be estimated as:  
\begin{equation}
N_\chi \simeq \frac{2}{3} \ln \left(\frac{\Gamma_{\phi\to a_b a_b}\theta^2}{\Gamma_{\chi \to H_u H_d}}\right) \simeq \frac{2}{3} \ln \left(\frac{3\, Y^4}{8 (\beta\,Z)^2}\,(\ln \vol)^4 \sqrt{\vol}\right).
\end{equation}
Finally, the reheating temperature from the volume decay is again given by (\ref{TrhVol}).
\end{itemize}
We have therefore obtained that $N_\phi$ and $N_\chi$ are different in each case but their sum, $(N_\phi + N_\chi)$, is the same in both cases, and looks like:
\begin{equation}
N_\phi+ N_\chi \simeq \frac23 \ln \left(\frac{H_{\rm inf}\,\theta^2}{\Gamma_{\chi \to H_u H_d}}\right).   
\end{equation}
As we shall see in the next section, this implies that both scenarios lead to the same number of e-foldings of inflation.

 \subsection{Inflationary parameters}

\begin{figure}[h]
\includegraphics[width=14cm]{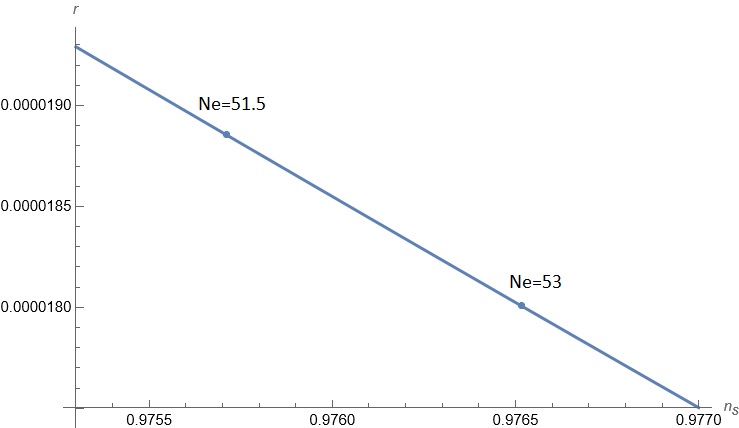}
\centering
\caption{Prediction of Loop Blow-up Inflation in the $(n_s,r)$-plane for a number of efoldings in the range $51.5\lesssim N_e\lesssim 53$.}
\label{Fig1}
\end{figure}

In order to get a prediction for the inflationary parameters, we have first to compute the number of efoldings of inflation $N_e$ based on the post-inflationary study performed in Sec.~\ref{sc:post-inflationary-dynamics}. The relevant formula, which takes into account the possibility to have two epochs of moduli domination with $N_\phi$ and $N_\chi$ efoldings respectively, reads \cite{Dutta:2014tya}:
\begin{equation}
\label{eq:N_e from reheating}
    N_e \simeq 57 +\frac{1}{4}\ln r -\frac{1}{4}\left(N_\phi + N_\chi\right) + \frac{1}{4} \ln \left(\frac{\rho_*}{\rho(t_{\rm end})}\right),
\end{equation}
where $\rho_*$ is the energy density at horizon exit. Since we are considering the potential \eqref{eq:fullpotential} which is rather flat during inflation, we shall assume  $\rho_* \simeq \rho(t_{\rm end})$ and neglect the last term in \eqref{eq:N_e from reheating}. The relation (\ref{eq:N_e from reheating}) for $N_e$ has to be combined with (\ref{eq:n_s corrected}), (\ref{r}) and (\ref{VolDetermination}) to obtain the predictions for $n_s$ and $r$ together with the value of the inflaton at horizon exit $\phi_*$ and the value of $\vol$ that allows to reproduce the observed amplitude of the density perturbations. For our illustrative parameter choice (\ref{Parameters}), we find:
\begin{equation}
\phi_* \simeq 0.06\, N_e^{7/22}\qquad \vol \simeq 1743\, N_e^{5/11}\qquad n_s \simeq 1 - \frac{1.25}{N_e} \qquad r\simeq \frac{0.004}{N_e^{15/11}}\,.
\end{equation}
Solving for $N_e$ in terms of $n_s$ and then substituting this result in the relation for $r$, we find the characteristic prediction of Loop Blow-up Inflation in the $(n_s,r)$-plane:
\begin{equation}
r \simeq 0.003 \left(1-n_s\right)^{15/11}\,.
\end{equation}
Fig. \ref{Fig1} shows this prediction for a number of efoldings in the range $49\lesssim N_e\lesssim 53$.

Using the analysis of Sec. \ref{sc:post-inflationary-dynamics}, we found numerically the values of $N_e$ and the resulting predictions for the main cosmological observables for all the scenarios studied above and for the following parameter choice:
\begin{equation}
Z=2\qquad \tilde c_{\rm loop} =1/(16\pi^2)\qquad Y=0.1 \qquad N_g=12\qquad g_*(T_{\rm rh})=106.75\,.
\end{equation}

\begin{itemize}
\item[\it I)] \textit{SM on D7s and inflaton wrapped by D7s:} 
\begin{equation}
N_\phi \simeq 1\qquad N_\chi \simeq 3\qquad N_e\simeq 53\qquad \phi_* \simeq 0.2 \qquad \vol \simeq 10616\,,
\end{equation}
with predictions:
\begin{equation}
n_s \simeq 0.9765 \qquad r\simeq 1.7\times 10^{-5}\qquad T_{\rm rh}\simeq 4\times 10^{10}\,{\rm GeV}\qquad \Delta N_{\rm eff}\simeq 0\,.
\label{Pred1}
\end{equation}

\item[\it II)] \textit{SM on D7s and inflaton not wrapped by any D7:} 
\begin{equation}
N_\phi \simeq 8\qquad N_\chi \simeq 0\qquad N_e\simeq 52\qquad \phi_* \simeq 0.2 \qquad \vol \simeq 10525\,,
\end{equation}
with predictions:
\begin{equation}
\label{eq:pred-2}
n_s \simeq 0.9761 \qquad r\simeq 1.7\times 10^{-5}\qquad T_{\rm rh}\simeq 3\times 10^{12}\,{\rm GeV}\qquad \Delta N_{\rm eff}\simeq 0.14\,.
\end{equation}

\item[\it III)] \textit{SM on D3s:} 
\begin{equation}
\left\{ \begin{array}{l} 
N_\phi \simeq 1\quad\,\, N_\chi \simeq 10.5 \quad\,\text{inflaton wrapped by D7s} \\
N_\phi \simeq 11\quad N_\chi \simeq 0.5 \quad\,\,\,\text{inflaton not wrapped by D7s}
\end{array} \right.
\quad\Rightarrow \quad N_\phi + N_\chi \simeq 11.5
\end{equation}
with:
\begin{equation}
N_e\simeq 51.5\qquad \phi_* \simeq 0.2 \qquad \vol \simeq 10447\,,
\end{equation}
and predictions:
\begin{equation}
n_s \simeq 0.9757\qquad r\simeq 1.8\times 10^{-5}    \qquad T_{\rm rh}\simeq 1\times 10^8\,{\rm GeV}\qquad\Delta N_{\rm eff}\simeq 0.36\,.
\label{Pred3}
\end{equation}
\end{itemize}
We see that the number of efoldings is around $N_e\simeq 52$-$53$, the reheating temperature is always large enough to allow for successful BBN, $10^8\,{\rm GeV}\lesssim T_{\rm rh}\lesssim 10^{12}\,{\rm GeV}$, and extra dark radiation can be compatible with present observational bounds. 

Moreover, the scalar spectral index goes from $n_s\simeq 0.9757$ to $n_s\simeq 0.9765$. These values have to be compared with CMB measurements which give \cite{Planck:2018vyg}: 
\begin{equation}
\label{eq:Planck value for n_s}
    n_s = 0.9665 \pm 0.0038 \quad\text{at}\,\,68\%\,\,\text{CL}\quad\text{for}\quad \Delta N_{\rm eff}=0\,.
\end{equation}
Note that the value \eqref{eq:Planck value for n_s} of $n_s$ has been inferred assuming the base-$\Lambda$CDM model with no extra dark radiation. Hence it should be compared only with the prediction (\ref{Pred1}), $n_s \simeq 0.9765$ and $\Delta N_{\rm eff}\simeq 0$, which is compatible with observations at $2.5\sigma$. This is already an acceptable matching with data, even if a better agreement could be achieved by including subleading perturbative corrections, as the ones discussed in Sec.~\ref{sc:more-models-ratios} or higher $\alpha'$ effects studied in~\cite{Cicoli:2023njy}.

The other predictions involve instead a non-zero amount of extra dark radiation, and so should be compared with CMB data fixing $\Delta N_{\rm eff}\simeq 0.14$ as in \eqref{eq:pred-2}, and $\Delta N_{\rm eff}\simeq 0.36$ as in \eqref{Pred3}. The Planck collaboration already performed the fit for $\Delta N_{\rm eff} = 0.39$ finding \cite{Planck:2015fie}:
\begin{equation}
\label{eq:n_s for higher Delta N_eff}
    n_s = 0.983 \pm 0.006\quad\text{at}\,\,68\%\,\,\text{CL}\quad \text{for} \quad  \Delta N_{\rm eff} = 0.39\,.
\end{equation}
Given that $\Delta N_{\rm eff} = 0.39$ is very similar to the value of extra dark radiation, $\Delta N_{\rm eff}\simeq 0.36$, of prediction (\ref{Pred3}), we can compare the value of $n_s$ in (\ref{Pred3}) with the one in (\ref{eq:n_s for higher Delta N_eff}), finding agreement within around $1.2\sigma$. To confront instead the predictions in \eqref{eq:pred-2} with observations, we should perform a fit similar to the one in \cite{Planck:2015fie} but fixing $\Delta N_{\rm eff}\simeq 0.14$. While this is beyond the scope of our paper, we can be very optimistic about this scenario. 
In fact, this case is middle-ground between \eqref{Pred1}, where our results are slightly higher than the corresponding value for $n_s$ resulting from CMB, and \eqref{Pred3}, where they fall slightly below. Moreover, we can use existing results for the extension of the base-$\Lambda$CDM model which includes $N_{\rm eff}$ as an additional parameter to fit cosmological data, leading to (c.f. Tables 4 and 5 of \cite{Planck:2018vyg}):
\begin{equation}
\label{eq:Planck value for n_s varying N_eff}
    n_s = 0.9589 \pm 0.0168\qquad\text{with }N_{\rm eff}=2.89^{+0.36}_{-0.38}\quad\text{at}\,\,95\%\,\,\text{CL}\,.
\end{equation}
It is then easy to see that the predictions for $n_s$ and $\Delta N_{\rm eff}$ in \eqref{eq:pred-2} agree with (\ref{eq:Planck value for n_s varying N_eff}) within around $2\sigma$. Therefore, we conclude that the scalar spectral index predicted by Loop Blow-Up Inflation is in good agreement with CMB observations.\footnote{Note that an even better agreement with cosmological data might be achieved in extensions of the $\Lambda$CDM model, like New Early Dark Energy \cite{Cruz:2022oqk}, which try to solve the $H_0$ tension.}

Finally, we stress that all scenarios lead to a similar value of $r$. Hence, Loop Blow-up Inflation predicts a tensor-to-scalar ratio of order:
\begin{equation}
r\simeq 2\times 10^{-5}\,, 
\label{rPrediction}
\end{equation}
which is within present observational bounds, $r<0.032$ at $98\%$ CL \cite{Tristram:2021tvh}, and it is much higher than the prediction of the original blow-up inflation model, $r \sim 10^{-10}$, \cite{Conlon:2005jm}.

\section{Discussion and conclusions}
\label{Concl}

The standard paradigm of slow-roll inflation involves potentials with an almost constant plateau. Interestingly, this picture can be reproduced in type IIb Calabi-Yau flux compactifications within the K\"ahler moduli sector, as mentioned and illustrated in different ways in \cite{Cicoli:2011zz, Burgess:2013sla, Burgess:2014tja, Burgess:2016owb, Cicoli:2023opf}. Let us summarize this general mechanism, commenting on how our new Loop Blow-up Inflation model compares with other models where inflation is driven by a K\"ahler modulus.

Given that $\mathcal{V}$ couples to all sources of energy due to the Weyl rescaling to go to 4d Einstein frame, which amounts to an overall $e^K = \mathcal{V}^{-2}$ multiplicative factor in front of the scalar potential, the volume mode is not a good inflaton candidate due to the impossibility to generate an inflaton-independent contribution to the scalar potential which is responsible for the inflationary plateau. Hence, the inflaton $\phi$ should be a direction orthogonal to $\mathcal{V}$. Moreover, since any contribution to the scalar potential is $\mathcal{V}$-dependent, a constant term requires the leading order dynamics to fix $\mathcal{V}$ while leaving $\phi$ unstabilized. This implies that $\phi$ should be a leading order flat direction that enjoys an approximate shift symmetry. This is the case for any modulus orthogonal to $\mathcal{V}$ in type IIb compactifications since the leading no-scale breaking effects are $\mathcal{O}(\alpha'^3)$ corrections which indeed depend just on $\mathcal{V}$. Balancing these effects against non-perturbative corrections for a diagonal del Pezzo divisor and different potential uplifting contributions (which also depend just on $\mathcal{V}$), can yield a dS minimum at exponentially large $\vol$ with $(h^{1,1}-2)$ flat directions. All of them can drive inflation once they are lifted at subleading order by additional quantum corrections. 

Hence the structure of the whole potential schematically looks like:
\begin{equation}
V_{\rm tot} (\mathcal{V},\tau_\phi)= V_{\rm lead}(\mathcal{V}) -  V_{\rm sub}(\mathcal{V},\tau_\phi)\,,
\end{equation}
where we ignored the $\mathcal{V}$-stabilizing blow-up mode, and for simplicity we focused just on one additional modulus $\tau_\phi$ that is a leading order flat direction since we assume the potential to have a hierarchical structure with $V_{\rm sub}(\mathcal{V},\tau_\phi) \ll V_{\rm lead}(\mathcal{V})$. Stabilizing the two fields gives:
\begin{equation}
\frac{\partial V_{\rm lead}}{\partial \mathcal{V}}(\langle\mathcal{V}\rangle) = 0\qquad\text{and}\qquad \frac{\partial V_{\rm sub}}{\partial \tau_\phi}(\langle\mathcal{V}\rangle,\langle\tau_\phi\rangle) = 0\,,
\end{equation}
with:\footnote{Note that $V_{\rm sub} \ll V_{\rm lead}$ refers to the inflationary regime, where $\tau_\phi$ is far away from its minimum value $\langle\tau_\phi\rangle$. There is hence no contradiction with the first equality in \eqref{vlevs}, which merely reflects the fine tuning of the cosmological constant in the post-inflationary vacuum.}
\begin{equation}
\label{vlevs}
V_{\rm lead}(\langle\mathcal{V}\rangle) = V_{\rm sub}(\langle\mathcal{V}\rangle,\langle\tau_\phi\rangle) \qquad\text{and}\qquad V_{\rm tot}(\langle\mathcal{V}\rangle,\langle\tau_\phi\rangle) = 0 \,,
\end{equation}
where we neglected the $\tau_\phi$-dependent shift of the volume minimum due to the large hierarchy between the two contributions to $V$. Setting $\mathcal{V}=\langle\mathcal{V}\rangle$, the potential thus becomes:
\begin{equation}
V_{\rm tot} (\langle\mathcal{V}\rangle,\tau_\phi) = V_{\rm sub}(\langle\mathcal{V}\rangle,\langle\tau_\phi\rangle)
 - V_{\rm sub}(\langle\mathcal{V}\rangle,\tau_\phi) = V_{\rm sub}(\langle\mathcal{V}\rangle,\langle\tau_\phi\rangle) \left[1 - \frac{V_{\rm sub}(\langle\mathcal{V}\rangle,\tau_\phi)}{V_{\rm sub}(\langle\mathcal{V}\rangle,\langle\tau_\phi\rangle)}\right].
\end{equation}
This potential takes a typical plateau-like form in terms of the canonically normalized inflaton $\phi$:
\begin{equation}
V= V_0\left[1-g(\phi)\right],
\label{Prototype}
\end{equation}
with:
\begin{equation}
V_0\equiv V_{\rm sub}(\langle\mathcal{V}\rangle,\langle\tau_\phi\rangle)   \qquad\text{and} \qquad g(\phi)\equiv \frac{V_{\rm sub}(\langle\mathcal{V}\rangle,\tau_\phi(\phi))}{V_{\rm sub}(\langle\mathcal{V}\rangle,\langle\tau_\phi\rangle)}\,,
\end{equation}
where the dependence on $\phi$ arises after replacing $\tau_\phi$ with $\phi$. Given that $\tau_\phi$ is a leading order flat direction, generically 
$V_{\rm sub}(\langle\mathcal{V}\rangle,\tau_\phi) \to 0$ for $\tau_\phi\to \infty$, or $V_{\rm sub}(\langle\mathcal{V}\rangle,\tau_\phi) \ll V_{\rm sub}(\langle\mathcal{V}\rangle,\langle\tau_\phi\rangle)$ for $\tau_\phi > \langle\tau_\phi\rangle$, guaranteeing the presence of an inflationary plateau for large values of $\tau_\phi$ where $g(\phi)\ll 1$ and $V\simeq V_0$. The exact expression of $g(\phi)$ depends on two features:
\begin{enumerate}
\item The origin (perturbative or non-perturbative) of the effects which generate $V_{\rm sub}(\langle\mathcal{V}\rangle,\tau_\phi)$:
\begin{itemize}
\item Perturbative effects are typically power-law and scale as:
\begin{equation}
V_{\rm sub}(\langle\mathcal{V}\rangle,\tau_\phi)\propto \frac{1}{\tau_\phi^p} \underset{\tau_\phi\to\infty}{\longrightarrow} 0 \qquad\text{for}\quad p>0\,,
\label{PertEffects}
\end{equation}
\item Non-perturbative effects are exponentially suppressed and behave as:
\begin{equation}
V_{\rm sub}(\langle\mathcal{V}\rangle,\tau_\phi)\propto e^{-k\tau_\phi} \underset{\tau_\phi\to\infty}{\longrightarrow} 0\qquad\text{for}\quad k>0\,.
\label{NPeffects}
\end{equation}
\end{itemize}

\item The topology of $\tau_\phi$ (a bulk or local cycle) which gives the relation between $\tau_\phi$ and $\phi$: 
\begin{itemize}
\item For a bulk modulus the canonical normalization introduces exponentials: 
\begin{equation}
    \tau_\phi = e^{\lambda \phi}\qquad\text{with}\qquad \lambda \sim \mathcal{O}(1)\,.
    \label{BulkCanNorm}
\end{equation}

\item For a local modulus the relation between $\tau_\phi$ and $\phi$ is power-law (see (\ref{eq:canonically normalized inflaton})):
\begin{equation}
\tau_\phi = \mu\,\mathcal{V}^{2/3}\,\phi^{4/3} \qquad\text{with}\qquad \mu\sim\mathcal{O}(1)\,.
\label{LocalCanNorm}
\end{equation}
\end{itemize}
\end{enumerate}
Together, these two features can give rise to different functional forms of $g(\phi)$. Assuming that $g(\phi)$ is rich enough to give a minimum at small field values, we focus just on its approximated expression in the inflationary region at large values of $\phi$, finding four inflationary scenarios which we name according to the features which determine $g(\phi)$: 
\begin{itemize}
\item \emph{Non-perturbative Blow-up Inflation}: if the potential arises from non-perturbative effects of the form (\ref{NPeffects}) for a local blow-up mode with canonical normalization given by (\ref{LocalCanNorm}), $g(\phi)$ in the inflationary region becomes:
\begin{equation}
 g(\phi)\propto \,e^{-k\mu\,\mathcal{V}^{2/3}\,\phi^{4/3}}\ll 1 \qquad\text{for}\qquad \phi > 0 \,.
\end{equation}
An inflationary model of this kind has been studied in~\cite{Conlon:2005jm, Bond:2006nc}, and subsequently in~\cite{Blumenhagen:2012ue} for the more involved case of a Wilson divisor (i.e. a rigid divisor with a Wilson line) with potential generated by poly-instantons (in this case a more appropriate name would therefore be \emph{Non-perturbative Wilson Inflation}). 
 
\item \emph{Non-perturbative Fibre Inflation}: if the potential is generated by non-perturbative effects of the form (\ref{NPeffects}) for a bulk fibration mode with canonical normalization given by (\ref{BulkCanNorm}), $g(\phi)$ becomes extremely small very quickly and in the inflationary region looks like:
\begin{equation}
 g(\phi)\propto e^{- k\, e^{\lambda\phi}}\ll 1 \qquad\text{for}\qquad \phi >0\,.
\end{equation}
This model has been developed in \cite{Cicoli:2011ct, Lust:2013kt} using poly-instanton effects for a fibre divisor.

\item \emph{Loop Fibre Inflation}: if the potential originates from perturbative corrections of the form (\ref{PertEffects}) for a bulk fibre divisor with canonical normalization given by (\ref{BulkCanNorm}), the dominant contribution to $g(\phi)$ in the inflationary region takes the form:
\begin{equation}
g(\phi) \propto e^{-p\lambda \phi}\ll 1\qquad\text{for}\qquad \phi>0\,.
\end{equation}
This model has been realized in \cite{Cicoli:2008gp, Cicoli:2016chb} which used a fibre divisor with potential generated by open string loops. The difference between the two models is just in the way the minimum is obtained: in \cite{Cicoli:2008gp} by balancing different loop contributions, whereas in \cite{Cicoli:2016chb} by balancing loops against higher $\alpha'$ effects. A similar model has been derived in \cite{Broy:2015zba} using a fibre divisor and a potential generated by higher $\alpha'$ effects (and so a more appropriate name for this model should be \emph{$\alpha'$ Fibre Inflation}).
 
\item \emph{Loop Blow-up Inflation}: this case corresponds to the new model developed in our paper where the inflationary potential is generated by perturbative effects of the form (\ref{PertEffects}) for a diagonal blow-up mode with canonical normalization given by (\ref{LocalCanNorm}). Hence, the form of $g(\phi)$ away from the minimum becomes:
\begin{equation}
g(\phi) \propto \frac{1}{\mathcal{V}^{2p/3}\,\phi^{4p/3}} \ll 1 \qquad\text{for}\qquad \phi \lesssim 1\,.
\end{equation}
In particular, in our model the potential is generated by string loops characterized by $p=1/2$. Substituting this value of $p$ in the form of $g(\phi)$, the potential (\ref{Prototype}) in the inflationary region reduces to:
\begin{equation}
V= V_0\left(1- \frac{c}{\mathcal{V}^{1/3}\phi^{2/3}}\right),
\label{Vmodel}
\end{equation}
which reproduces exactly the potential (\ref{eq:fullpotential}) of our model after identifying $c=c_{\rm loop}\,\sigma_\phi/\beta$. This potential can naturally drive inflation since it can be approximated as a constant plateau, $V\simeq V_0$, for $\phi\lesssim 1$ thanks to the $\vol^{-1/3}$ suppression factor.
\end{itemize}
The main concern regarding Loop Blow-up Inflation, as already pointed out in \cite{Cicoli:2011zz}, is that, in order to get accelerated expansion with $\epsilon <1$, $\phi$ might have to be pushed to $\mathcal{O}(1)$ values which correspond to $\tau_\phi\sim\mathcal{V}^{2/3}$, as can be seen from (\ref{LocalCanNorm}). In this region of moduli space the blow-up mode $\tau_\phi$ becomes as large as the overall volume, and so we are close to the walls of the K\"ahler cone where the EFT could be out of control. However in this paper we have performed a detailed analysis of the inflationary and post-inflationary dynamics showing that phenomenologically viable slow-roll can be achieved far enough from the boundaries of the K\"ahler cone since horizon exit occurs at $\phi_*\simeq 0.2$ where the EFT is still under control. We stress that the smallness of the coefficient of the loop correction, which follows both in analogy to the familiar 4d loop suppression factor $1/(16\pi^2)$ and by considering explicit torus orbifold results,
is crucial to achieve this conclusion. 

Let us also point out that all K\"ahler moduli inflation models built so far feature an exponential potential in terms of the canonical inflaton $\phi$, and so the potential of our model (\ref{Vmodel}) represents the first example in this class of constructions of a power-law inflationary potential. A crucial ingredient to obtain such a potential is the presence of string loop corrections to the K\"ahler potential which we argued to be inevitable. These perturbative effects can be subdominant with respect to non-perturbative corrections close to the minimum for $\tau_\phi$ but then quickly come to dominate the scalar potential when $\tau_\phi$ is displaced away from the minimum. We estimated that, in order to reproduce the original model of blow-up inflation driven by non-perturbative effects \cite{Conlon:2005jm}, the coefficient of the loop corrections should be tiny, $\cloop \ll 10^{-6}$. Hence, whenever $c_{\rm loop}\gtrsim 10^{-6}$, loop effects are large and they are the leading no-scale breaking effects along $\tau_\phi$ for large field values. 

Focusing on the natural regime where $c_{\rm loop}\gtrsim 10^{-6}$, we studied in depth the simplest realization of Loop Blow-up Inflation which involves just one additional blow-up mode $\tau_\phi$ and the leading loop corrections. In particular, we derived the predictions for the main cosmological observables in terms of the underlying parameters and the number of efoldings $N_e$. In turn, we determined $N_e$ from studying the rich post-inflationary evolution of our model which features in general a non-standard thermal history with epochs of moduli domination. Depending on the microscopic brane setup and realization of the SM, we found $N_e$ in the range $51.5\lesssim N_e\lesssim 53$ and a rather high reheating temperature, $10^8\,{\rm GeV}\lesssim T_{\rm rh}\lesssim 10^{12}\,{\rm GeV}$. In order to reproduce the observed amplitude of primordial density perturbations, the CY volume has to be of order $\vol\simeq 10^4$ and horizon exit occurs at $\phi_*\simeq 0.2$. Moreover, extra dark radiation due to the production of ultra-light bulk axions from moduli decays can be within observational bounds and the scalar spectral index is in agreement with CMB data. Finally, Loop Blow-up Inflation predicts the relation $r\simeq 0.003\left(1-n_s\right)^{15/11}$ between the spectral index and the tensor-to-scalar ratio, implying the numerical value $r\simeq 2\times 10^{-5}$.

Lastly, we also discussed the effect of loop corrections subleading in the parameter $\tau_\phi/{\cal V}^{2/3}$. These might become important as we get close to the boundaries of the K\"ahler cone. We argued that these corrections could realize a whole class of inflationary models at large values of the inflaton in a potentially controlled manner. In order to explore these effects more concretely, it would however be necessary to perform explicit loop calculations in a specific CY geometry, a task that is notably complex and challenging. Moreover, as a future direction of work along the lines of \cite{Cicoli:2023njy}, it would be interesting to include additional perturbative corrections like higher $F$-term $\alpha'^3$ effects \cite{Ciupke:2015msa}.

Finally, let us stress that we treated $c_{\rm loop}$ as a phenomenological parameter since an explicit top-down computation of the coefficient of the loop corrections to the K\"ahler potential is a technically very challenging task.
The sign and magnitude of $c_{\rm loop}$ are crucial for the realization of our inflationary scenario.
In analogy with explicit toroidal computations, $c_{\rm loop}$ is in general expected to be a function of the complex structure moduli, and so it might enjoy a certain degree of tuning freedom in the type IIb flux landscape. 
Moreover, we expect that for the simplest blow-up geometries progress can be made by an explicit computation of $\cloop$ in the regime of large volume and small blow-up cycle. In this regime, the geometry near the blow-up can be approximated by a non-compact Calabi-Yau.
For example, one could analyse the blown up $\mathbb{C}^3/\mathbb{Z}_3$ where an explicit Ricci-flat metric is known~\cite{Calabi1979,Higashijima:2001fp,Higashijima:2001vk,Higashijima:2002px,Ganor:2002ae,GrootNibbelink:2007lua}. While an explicit loop calculation appears feasible in such a relatively simple geometry, this is beyond the scope of our paper.

\section*{Acknowledgements}

We would like to thank Joe Conlon, Anshuman Maharana, Francisco G. Pedro, Fernando Quevedo, Filippo Sala, Simon Schreyer, Gian Paolo Vacca and Roberto Valandro for useful conversations. The work of AH is supported by the Deutsche Forschungsgemeinschaft (DFG, German Research Foundation) under Germany’s Excellence Strategy EXC 2181/1 - 390900948 (the Heidelberg STRUCTURES Excellence Cluster).
RK is  supported by the International Max Planck Research School for Precision Tests of Fundamental Symmetries (IMPRS-PTFS). The work of LB and MC contributes to the COST Action COSMIC WISPers CA21106, supported by COST (European Cooperation in Science and Technology).

\appendix
\section{Comment on loop corrections}
\label{app:loop-notation}

The \K potential \eqref{eq:kahler-potential} is subject to loop corrections. These have been estimated using 10d EFT arguments \cite{vonGersdorff:2005bf} and explicitly calculated for torus-based geometries in a string one-loop analysis \cite{Berg:2005ja}. The torus-orbifold result is commonly written as a sum of Kaluza-Klein (KK) and winding (W) corrections \cite{Berg:2007wt}:
\begin{equation}
    \delta K_{(\gs)} = \delta K^\KK_{(\gs)}+\delta K^\W_{(\gs)}\,.
\end{equation}
Extrapolating the torus result, it was conjectured in \cite{Berg:2007wt} that the corrections on a generic Calabi-Yau take the form
\begin{equation}
    \delta K^\KK_{(\gs)} \simeq \sum\limits_{i} C_i^\KK~\frac{\gs {\cal T}^i(t^a) }{\vol}\,\,\,,\qquad
    \qquad \delta K^\W_{(\gs)}\simeq \sum\limits_{i}C_i^\W~\frac{ 1 }{{\cal I}^i(t^a)\vol}\,,
\end{equation}
where the coefficients $C_i^\KK$ and $C_i^\W$ are unknown functions of the complex structure moduli and are expected to be suppressed by $\pi$ factors (for estimates cf. the last paragraph on p.~41 of \cite{Gao:2022uop}). The functions ${\cal T}^i$ and ${\cal I}^i$ were conjectured in \cite{Berg:2007wt} to be linear in the 2-cycle volumes $t^i$. Later on, it was argued in \cite{Gao:2022uop} that more general functional forms arise and that one should only expect ${\cal T}^i$ and ${\cal I}^i$
to be homogeneous functions of the 2-cycle volumes of degree 1.

Because $\delta K^\KK_{(\gs)}$ is of degree $-2$ in the 2-cycle volumes, the correction in the scalar potential $\delta V_{(\gs)}^\KK$ has an `extended no-scale structure' \cite{Berg:2005yu,vonGersdorff:2005bf,Berg:2005ja,Berg:2007wt,Cicoli:2007xp}.
In the end, using the linearity assumption for ${\cal T}^i$, ${\cal I}^i$, the corrections to the scalar potential $\delta V_{(\gs)}$ read \cite{Cicoli:2007xp,Cicoli:2008va}:
\begin{equation}
    \label{eq:V-corrections}
    \delta V_{(\gs)} =  \frac{\ W_0^2}{\vol^2} \left( (\gs C_i^\KK )^2 K^{\text{tree}}_{ii} -2 \delta K^\W_{(\gs)} \right),
\end{equation}
where the tree-level K\"ahler potential is $K^{\text{tree}}=-2\ln \vol$ and we have omitted the prefactor $\hat{V}$ defined in (\ref{eq:LVS-potential-prefactors}). Using a form of the volume $\vol$ as in \eqref{eq:Swiss-volume}, we find to leading order
\begin{equation}
    \label{eq:final-loop-corrections}
    \delta V_{(\gs)} \simeq \frac{\ W_0^2}{\vol^3}\frac{c_\text{loop}}{\vol^{1/3}}\left( \frac{\vol^{1/3}}{\sqrt{\tau_i}} 
    +\O(1) + \O \left(\frac{\sqrt{\tau_i}}{\vol^{1/3}}\right)\right), \quad c_\text{loop}\simeq\begin{cases}C_i^\W\\(\gs C_i^\KK )^2
    \end{cases}\!\!\!\!,
\end{equation}
where we introduced the coefficient $\cloop$ to remain agnostic about the origin of the loop corrections. Ref. \cite{Cicoli:2007xp} provided a field-theory interpretation of the result (\ref{eq:final-loop-corrections}) matching it with the one-loop Coleman-Weinberg potential which in supergravity reads
\begin{equation}
V^\CW_{\rm 1-loop}\simeq \frac{1}{16\pi^2}\,\Lambda^2\,{\rm Str} M^2\,,
\label{CW}
\end{equation}
where the EFT cut-off $\Lambda$ can be identified with the mass of Kaluza-Klein replicas of open string modes living on D7-branes wrapped around different 4-cycles
\begin{equation}
\Lambda\simeq\begin{cases}\Lambda_i\simeq \frac{1}{\tau_i^{1/4}\sqrt{\vol}}\qquad\text{for D7s on $\tau_i$ with $i=\phi,s$} \\ 
\Lambda_b \simeq \frac{1}{\vol^{2/3}}\qquad\quad\text{for D7s on $\tau_b$}\,\,.
    \end{cases}   
\end{equation}
Using these cutoff scales in (\ref{CW}) together with the estimate ${\rm Str}M^2\simeq m_{3/2}^2\simeq W_0^2/\vol^2$, one can reproduce the scaling of the first two terms in (\ref{eq:final-loop-corrections}), justifying the smallness of the coefficient $c_{\rm loop}$ which is expected to scale as $c_{\rm loop}\simeq 1/(16\pi^2)$.\footnote{Following \cite{Berg:2007wt}, ref. \cite{Cicoli:2007xp} proposed to match also the third therm in (\ref{eq:final-loop-corrections}) exploiting flux-dependent correction to $\Lambda_i$.}

The comparison of field-theoretic and string one-loop logic has been perfected in \cite{Gao:2022uop}. One conclusion is that corrections with the parametric form of $\delta K^W_{(g_s)}$ do not only arise from fields on D7 brane intersections \cite{Berg:2007wt,Cicoli:2007xp} but also from closed strings or, equivalently, 10d fields. From this perspective, even in the absence of D7-branes wrapping the $\tau_i$ cycle, the cut-off scale $\Lambda_i$ should be identified with the mass of Kaluza-Klein modes with wavelength $\sim\tau_i^{1/4}$, which are closed string states.
This implies that $c_{\rm loop}$ unavoidably includes a non-zero piece without $g_s$ suppression. Moreover, for the case of a blow-up cycle, a derivation of the leading term in \eqref{eq:final-loop-corrections} was provided in \cite{Gao:2022uop}, in agreement with the leading term following from the conjecture of \cite{Berg:2007wt}. Finally, note that one should really read 
\eqref{eq:final-loop-corrections} with the replacement
\begin{equation}
\left( \frac{\vol^{1/3}}{\sqrt{\tau_i}} +
\O(1)+\O \left(\frac{\sqrt{\tau_i}}{\vol^{1/3}}\right)\right) \quad\rightarrow \quad f\left(\frac{\vol^{1/3}}{\sqrt{\tau_i}}\right)\,,
\end{equation}
where $f$ encodes information from the unknown functions ${\cal T}^i$ and ${\cal I}^i$.

A key ingredient of LVS constructions is the presence of 3-form fluxes. Let us therefore comment on their effect on the loop corrections that our inflationary model is based on. At large volume, such terms are expected to be subleading for the following reason. First, the superpotential is subject to the standard ${\cal N}=1$ non-renormalization theorem, and so it is not corrected at perturbative level. Second, the loop corrections to the K\"ahler potential can be understood as an infinite sum of one-loop corrections from all KK modes propagating in the compact CY orientifold geometry. At leading order, this KK mode spectrum depends only on the (flux-less) orientifold geometry. This geometry leads to a spectrum that displays only ${\cal N}=1$ SUSY and hence produces a non-zero loop correction. Of course, fluxes disturb this geometry through warping and also affect the KK mode spectrum directly since the 10d action involves vertices (3-point and higher) between $B_2/C_2$ and other fields. However, both of these effects are proportional to the 3-form field strength, and so vanish as the fluxes become more and more dilute at large volume. The resulting effect on the loop correction is hence more strongly volume suppressed than the loop effect resulting merely from the propagation of 10d fields in the unperturbed orientifold geometry.

We can be more quantitative by noting that, in the absence of fluxes, the KK mass-squared scales as $M_{{\rm KK},0}^2\sim \left(M_s/\ell\right)^2$, where $M_s$ is the string scale and $\ell$ is a typical CY radius in string units. This can be viewed as an energetic effect associated with the excitation of a KK mode deformation of the geometry. If a 3-form flux with integer flux number $N$ is present, a competing effect of order $\delta M_{\rm KK}^2\sim \left(N\,M_s/\ell^3\right)^2$ is expected to arise. After Weyl rescaling to 4d Einstein frame, i.e. writing $M_s\sim M_p/\sqrt{\mathcal{V}}$ where $\mathcal{V}$ is the CY volume in string units, and considering $N\sim W_0$, we obtain:
\begin{equation}
M_{\rm KK}^2\sim \frac{M_p^2}{\mathcal{V}\, \ell^2} \left(1+ \frac{W_0^2}{\ell^4}\right)
\label{KKmass}
\end{equation}
which agrees with the estimate of App. D of \cite{Berg:2007wt}. Note that the correction in (\ref{KKmass}) is precisely the flux stabilization scale of the zero-modes of the complex structure moduli, $W_0^2/{\cal V}^2$, taking $\ell^6\sim \mathcal{V}$. Given that 3-form fluxes are known to lift only the complex structure moduli, it might be that the correction in (\ref{KKmass}) is absent for the KK modes of the K\"ahler moduli or that, if present, it introduces a dependence just on the complex structure moduli, i.e. that $\ell$ is a 3-cycle radius. In this case these corrections would not induce any dependence on the inflaton $\tau_\phi$ which, being a K\"ahler modulus, is a 4-cycle radius. However, even in the case where $\ell^4= g_s\,\tau_\phi$, such a correction would be harmless for the inflationary dynamics. In fact, at the minimum $\tau_\phi$ is of order $g_s^{-1}$, and so around the minimum $\ell\sim \mathcal{O}(1)$. However inflation takes place in the region in moduli space far away from the minimum where $\tau_\phi\gg g_s^{-1}$, resulting in $\ell\gg 1$ which makes the correction in (\ref{KKmass}) very suppressed for $W_0\sim\mathcal{O}(1)$.

Finally, we note that in the presence of fluxes and O-planes/D-branes, one generally also has to deal with warping and a non-trivial dilaton profile. This induces further $\alpha'$-suppressed corrections \cite{Grimm:2013gma, Minasian:2015bxa}. We do not expect them to affect our model at leading order and we do not discuss them in this work.

\bibliographystyle{JHEP}
\bibliography{references}
\end{document}